\documentclass[prl,twocolumn,amsmath,amssymb,showpacs,superscriptaddress]{revtex4-2}
\pdfoutput=1
\UseRawInputEncoding
\usepackage{graphicx}
\usepackage{epstopdf}
\usepackage{dcolumn}
\usepackage{bm}
\usepackage[FIGTOPCAP]{subfigure}
\usepackage[usenames]{color}
\usepackage{mathtools}
\usepackage[all]{xy}

\def \bea{\begin{eqnarray}}
	\def \eea{\end{eqnarray}}

\definecolor{mygray}{gray}{0.8}
\definecolor{plum}{rgb}{.5,0,1}
\begin{document}
	\title{Play. Pause. Rewind. Measuring local entropy production and extractable work in active matter}
	\author{Sunghan Ro}
	\thanks{Equal contribution.}
	\affiliation{Department of Physics, Technion-Israel Institute of Technology, Haifa 3200003, Israel}
	
	\author{Buming Guo}
	\thanks{Equal contribution.}
	\affiliation{Center for Soft Matter Research, Department of Physics, New York University, New York 10003, USA}
	
	\author{Aaron Shih}
    \affiliation{Courant Institute of Mathematical Sciences, New York University, New York 10003, USA}
	\affiliation{Department of Chemical Engineering and Materials Science, University of Minnesota, Minneapolis, Minnesota 55455, USA}
	
	\author{Trung V. Phan}
	\affiliation{Department of Physics, Princeton University, Princeton 08544, New Jersey, USA}
	
	\author{Robert H. Austin}
	\affiliation{Department of Physics, Princeton University, Princeton 08544, New Jersey, USA}
	
	\author{Dov Levine}
	\email{dovlevine19@gmail.com}
	\affiliation{Department of Physics, Technion-Israel Institute of Technology, Haifa 3200003, Israel}
	
	\author{Paul M. Chaikin}
	\email{chaikin@nyu.edu}
	\affiliation{Center for Soft Matter Research, Department of Physics, New York University, New York 10003, USA}
	
	\author{Stefano Martiniani}
	\email{sm7683@nyu.edu}
    \affiliation{Center for Soft Matter Research, Department of Physics, New York University, New York 10003, USA}
    \affiliation{Courant Institute of Mathematical Sciences, New York University, New York 10003, USA}
    \affiliation{Department of Chemical Engineering and Materials Science, University of Minnesota, Minneapolis, Minnesota 55455, USA}
    \affiliation{Simons Center for Computational Physical Chemistry, Department of Chemistry, New York University, New York 10003, USA}

\begin{abstract}
	Time-reversal symmetry breaking and entropy production are universal features of nonequilibrium phenomena. Despite its importance in the physics of active and living systems, the entropy production of systems with many degrees of freedom has remained of little practical significance because the high-dimensionality of their state space makes it difficult to measure. Here we introduce a local measure of entropy production and a numerical protocol to estimate it. We establish a connection between the entropy production and extractability of work in a given region of the system and show how this quantity depends crucially on the degrees of freedom being tracked. We validate our approach in theory, simulation, and experiments by considering systems of active Brownian particles undergoing motility induced phase separation, as well as active Brownian particles and E. Coli in a rectifying device in which the time-reversal asymmetry of the particle dynamics couples to spatial asymmetry to reveal its effects on a macroscopic scale.
\end{abstract}

\maketitle

Time reversal symmetry breaking (TRSB) in active matter systems arises from the self-propelled motion of individual particles, driven by the continuous injection of energy into the system~\cite{marchetti2013hydrodynamics,fodor2021irreversibility,o2022time,tailleur2008statistical}. TRSB is linked to the emergence of steady-state currents in phase-space, a clear signature of a system's departure from equilibrium, which can be quantified in terms of the global rate of entropy production~\cite{seifert_stochastic_2012,van2015ensemble,mandal2017entropy,shankar2018hidden,gnesotto2018broken,kurchan1998fluctuation,lebowitz1999gallavotti,seifert2010fluctuation,jarzynski2011equalities}.

Notwithstanding its importance, global entropy production has not yet found a place in the routine characterization of experimental many-body nonequilibrium systems. The reasons for this are twofold. First, the high-dimensional nature of phase-space has hindered the estimation of entropy production beyond low-dimensional systems~\cite{andrieux_entropy_2007,li2019quantifying,manzano2021thermodynamics,skinner2021improved,pietzonka2017entropy}. Although multiple approaches have been adopted to mitigate this problem from a variety of perspectives~\cite{liphardt2002equilibrium,roldan2012entropy,battle2016broken,kim2020learning}, an adequate solution remains elusive. 

The second, and arguably more fundamental reason, is that global entropy production is a single number which, as such, cannot provide insight into the complex patterns of TRSB events occurring in many-body systems. Crucial theoretical insight into this problem came from Nardini et al.~\cite{nardini2017entropy} who, moving beyond the estimation of the global quantity, proposed a spatially local decomposition of entropy production. To this end, they analyzed stochastic field theories of active model systems undergoing motility-induced phase separation (MIPS) showing that at a coarse-grained level TRSB events are pronounced at the interfaces while reversibility is partially recovered in bulk, thus casting light on where the system's departure from equilibrium is more prominent. However, dynamics of local degrees of freedom obtained by coarse-graining are in many cases non-Markovian~\cite{martinez2019inferring,skinner2021improved}, which makes the accurate estimation of the local entropy production difficult.

To overcome this problem, we introduce a universal information-theoretic measure applicable to non-Markovian dynamics, and a numerical protocol to estimate it. We then apply this protocol to measure a ``local'' entropy production in several theoretical and experimental systems. We show that the measured local entropy production provides comprehensive information about the nature of TRSB events in the system, such as their characteristic length scales, as well as whether any work can be extracted when an external mechanism is weakly coupled to the tracked degrees of freedom.

\begin{figure*} [t!]
	\center
	\includegraphics[width=1.0\linewidth]{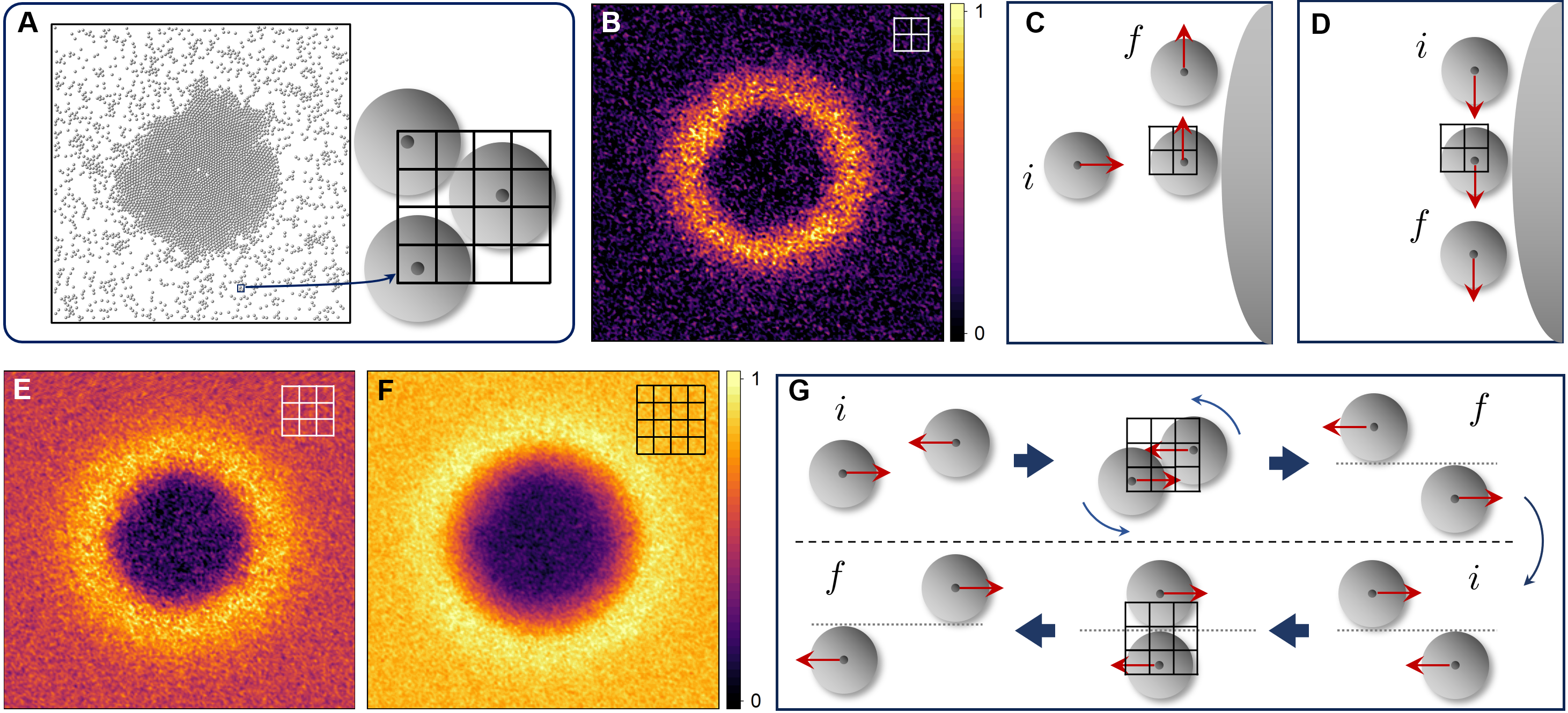}
	\caption{ Normalized local entropy production rate, EP, for a MIPS cluster.  (A) Active Brownian Particle, ABP, presenting MIPS in simulation,  (B) EP with $2\times 2$ blocks. (C) Typical motion of an ABP approaching an interface. (D) Typical motion of an ABP propelling parallel to the interface. (E)-(F) EP with $3 \times 3$ and $4 \times 4$ blocks, that capture progressively more degrees of freedom and EP in the dilute phase. (G) TRSB of two ABPs colliding with each other. Normalization factors for EPs are chosen so that the fraction of samples with EPs greater than these factors is less than $10^{-3}$. Specifically, we set the factors as (B) $2.3 \times 10^{-3}$, (E) $7.8 \times 10^{-3}$, and (F) $2.7 \times 10^{-2}$.
	}
	\label{Fig:ABP}
\end{figure*}

{\it Local entropy production rate}.---
We start by introducing the concept of local entropy production and demonstrating its usefulness for characterizing nonequilibrium features in driven many-body systems. 
To provide a concrete picture, we consider an example of active Brownian particles (ABPs) with excluded volume interactions, undergoing motility-induced phase separation (MIPS). The ABPs propel themselves at a constant speed, while the propulsion direction diffuses randomly. Remarkably, for high enough speed, ABPs accumulate with each other and form a dense cluster even when the particles do not attract each other, ultimately giving rise to MIPS (see Fig.~\ref{Fig:ABP}A for a snapshot). 

Although MIPS is indeed a non-equilibrium phenomenon, phase separation itself can also be observed in equilibrium. This leads to the question of whether MIPS present genuinely non-equilibrium features or not. Among many such features suggested so far, we focus on those associated with entropy production~\cite{fodor2016far,nardini2017entropy,martin2021statistical,fodor2021irreversibility}. Since ABPs are clearly driven out of equilibrium, entropy is constantly produced by the system. To measure entropy production (EP), we track the system configuration $X(t)$, consisting of both positions and orientations, over a given observation interval $\tau$, yielding the global state trajectory ${\bf X}$. 
Due to lack of TRS in the system dynamics, we expect the time-reversed trajectory ${\bf X}^\mathrm{R}$ obtained by playing ${\bf X}$ backward to be distinguishable from the original forward trajectory. Accordingly, the global EP can be quantified by the Kullback-Leibler divergence between the forward and the backward realizations of the trajectories: $\sigma = \tau^{-1} \langle \ln ({\cal P}[{\bf X}] / {\cal P}[{\bf X}^\mathrm{R}] \rangle$, where ${\cal P}[{\bf X}]$ is the probability to observe ${\bf X}$ in the forward dynamics and the angled brackets denote the average over ${\cal P}[{\bf X}]$.

The global EP is a single positive number whose non-zero value only signifies that the system is out of equilibrium. This obscures the spatial inhomogeneity of the system: for example, we expect distinctive patterns of non-equilibrium particle motion in the dense phase, in the dilute phase, and at the interface between them. 
These spatial patterns are, however, completely erased in the global EP. To recover this information, we turn to measuring the local EP by overlaying an $L \times L$ square grid over the system and specifying local states by observing the particle occupancy in a small block chosen from the grid.
In doing so, we set the grid size so that only one particle can fit in a $2 \times 2$ block to ensure fine enough spatial resolution, precluding multiple occupancy and ambiguity in the particle motion. 
Then, we first associate each grid point with the state of a $2 \times 2$ block which results in one of five possible configurations: either empty or one of the four sites being occupied. We denote this local state at time $t$ as $\chi_i(t)$ where the index $i = (x, y)$ with $x, y \in [1:L]$ specifies the position of the block. Again, by tracking $\chi_i(t)$ during a time interval $\tau$, we obtain a local trajectory $\boldsymbol{\chi}_i$, with which we define the local EP
\begin{equation} \label{Eq:local_EP}
    \sigma_i = \frac{1}{\tau} \left\langle \ln ({\cal P}[\boldsymbol{\chi}_i] / {\cal P}[\boldsymbol{\chi}_i^\mathrm{R}] \right\rangle~,
\end{equation}
where $\boldsymbol{\chi}_i^\mathrm{R}$ is the time-reversed realization of the local trajectory and ${\cal P}[\boldsymbol{\chi}_i]$ refers to the PDF of the local trajectory. 

Using the measure of EP that we will introduce presently, we evaluated $\sigma_i$ for a range of positions $i$, yielding a spatially resolved map that cannot be captured by the global EP. 
In Fig.~\ref{Fig:ABP}B we present a map of local EP obtained from the local states defined by the occupancy of $2 \times 2$ blocks. The result shows that local EP is concentrated at the interface between the phases formed by MIPS. To understand this pattern, it is useful to consider the typical motion of a particle from an initial state $i$ to a  final state $f$ near an interface, as depicted in Fig.~\ref{Fig:ABP}C. If a particle approaches the interface from a normal direction, it is likely to stay at the contact location while propelling toward the droplet center. Once the propulsion direction reorients itself (via rotational diffusion) so that it becomes roughly parallel to the interface, the particle will move away from the contact location. 
The time-reversed realization of this event is highly unlikely, as a particle moving along the interface is likely to pass by without any abrupt turns, as depicted in Fig.~\ref{Fig:ABP}D. This imbalance leads to high local EP measured at the interface. Note that this result is in agreement with the field-theoretic predictions~\cite{wittkowski2014scalar, nardini2017entropy,martin2021statistical}, which we reproduce for active model B in SI.

While the spatial pattern captured by the local EP of $2 \times 2$ blocks is remarkable, it can only detect TRSB events that depend on the dynamics of a single particle, ignoring those TRSB events that depend explicitly on the dynamics of multiple particles. Accordingly, we explore the effect of computing local EP from the local states obtained using larger grid blocks, such as $3 \times 3$ or $4 \times 4$ blocks, which can track two or more particles simultaneously (see Fig.~\ref{Fig:ABP}A). The local EP obtained with $3 \times 3$ blocks shown in Fig.~\ref{Fig:ABP}E still presents the highest EP at the interface but there is also considerable entropy production in the dilute phase. This additional EP captured from the dilute phase originates from irreversible collision events between two active particles, depicted in the upper portion of Fig.~\ref{Fig:ABP}G. In detail, due to the persistence of the propulsion, colliding ABPs slide along each other until their propulsion directions become perpendicular to the line connecting their centers, at which point the particles go past each other. 
If we reverse the propulsion directions of the particles from the final state, we will not observe the time-reversed realization of the forward trajectory as depicted in the lower portion of Fig.~\ref{Fig:ABP}G. 
Further enlarging the block size to $4 \times 4$ reveals even more EP in the dilute phase and a marginal increase of EP in the dense phase, as shown in Fig.~\ref{Fig:ABP}F.

Intriguingly, comparisons between the local EPs obtained with various definitions of local states allows us to characterize the nature of TRSB events occurring at various locations in the system. At the interface, TRSB events are due to collision between particles and the interface and occur on the scale of a particle radius, as indicated by the fact that they could be captured with $2 \times 2$ blocks. Meanwhile, there are plenty of TRSB two-particle collision events occurring in the dilute phase whose length scale is equal to the particle diameter, and thus requires larger blocks to capture. These examples demonstrates that local EPs measured with various choices of local states may enable characterization of non-equilibrium events occurring in a variety of systems.

In SI, we further provide local EP measurement results obtained by further considering particle orientations. In this case, we observe that tracking $1\times 1$ blocks is sufficient to capture TRSB events at the interface, while tracking $2 \times 2$ blocks is sufficient to highlight the distinction between the dense and the dilute phases. We also report that similar results are reproduced with active particles strictly on lattice.

{\it Measure of entropy production rate}.---
We now proceed to the problem of how to actually measure local EP as defined in Eq.~\eqref{Eq:local_EP}. As it is evident from the equation, direct evaluation would requires the PDF of the local state trajectories ${\cal P}[\boldsymbol{\chi}_i]$. Obtaining ${\cal P}[\boldsymbol{\chi}_i]$ can, however, be a daunting task since the local dynamics exhibit a finite memory time due to interactions between the local degrees of freedom and their environment~\cite{ziv_measure_1993}.

To address this problem, we propose a measure based on the information-theoretic quantity known as {\it cross-parsing complexity}, ${\cal C}$, introduced by Ziv and Merhav~\cite{ziv_measure_1993,wyner_role_1998-1}. To compute ${\cal C}$ we must consider two sequences that we refer to as the {\it sample} sequence and the {\it database}, respectively. We proceed by compressing the sample sequence by comparing it to the database, as follows: Starting at position $i=0$ of the sample sequence, we look for the longest common word (or factor) anywhere in the database, yielding a word of length $\ell_1$. We then repeat this process starting from position $i=\ell_1$ of the sample sequence to find the next longest match of size $\ell_2$, and so on until we get to the end of the sample sequence. Assuming that the database and the sample sequence have lengths $M$ and $N$, respectively, we will have decomposed the sequence in $\mathcal{C}$ longest common factors, such that $N = \sum_{j=1}^{\mathcal{C}} \ell_j$. Clearly, the sample sequence can be encoded in roughly $N^{-1} \mathcal{C} \log M + N^{-1} \sum_{j=1}^{\mathcal{C}} \log \ell_j \leq   N^{-1} \mathcal{C} \log M + N^{-1} \mathcal{C} \log \langle \ell \rangle$ bits per character, where we arrived at the inequality by convexity of the $\log$. Substituting $\langle \ell \rangle = N/\mathcal{C}$ and using the fact that $\langle \ell \rangle \sim O(\log N)$, one can easily verify the cross-entropy is bounded as
\begin{equation}
\label{cross-entropy-bound}
    H(\mathrm{sample}, \mathrm{database}) \leq  \mathcal{C} \frac{\log M}{N}.
\end{equation}
where $H(p, q) = - \sum_i p_i \log q_i$ is the cross entropy for the distributions $p, q$ (when $p=q$ this is simply the entropy $H(p)$). Sample sequences that are similar to the database have small cross-parsing complexities, ${\cal C}$, while sequences that are dissimilar to the database have large $\mathcal{C}$. From Eq.~\ref{cross-entropy-bound} and the fact that the KL-divergence of the sample from the database is equal to their cross entropy minus the entropy of the sequence i.e., $\mathrm{KL}(\mathrm{sample}||\mathrm{database}) = H(\mathrm{sample}, \mathrm{database}) - H(\mathrm{sample})$, we find that the difference of the cross-parsing complexity between independently sampled forward and backward sequences, ${\cal C}({\boldsymbol{\chi}_i^\prime}||{\boldsymbol{\chi}_i}^\mathrm{R})$, and between independently sampled forward sequences, ${\cal C}({\boldsymbol{\chi}_i^\prime}||{\boldsymbol{\chi}_i})$, gives the EP
\begin{equation} \label{eq:measure}
    \tilde{\sigma}_i = \lim_{N \to \infty}  \frac{\ln N}{N} \left[ {\cal C}({\boldsymbol{\chi}_i^\prime}||{\boldsymbol{\chi}_i}^\mathrm{R}) - {\cal C}({\boldsymbol{\chi}_i^\prime}||{\boldsymbol{\chi}_i}) \right]~,
\end{equation}
which is the measure (estimator) that we propose, where we assumed $M=N$. In SI, we show that our estimator exhibits significantly faster convergence than the original approach by Ziv and Merhav~\cite{ziv_measure_1993}, as well as the subsequent refinement by Rold\'{a}n and Parrondo~\cite{roldan2010estimating, roldan2012entropy}. Note that ${\cal C}$ can be obtained efficiently in ${\cal O}(N)$ time using suffix arrays \cite{karkkainen2013linear}. More importantly, Eq.~\eqref{eq:measure} can be computed without prior knowledge on the memory time. Finally, we note that Eq.~\eqref{eq:measure} is defined for a discrete sequence, so we must subsample a continuous time trajectory by a time interval $\Delta t$ yielding a discrete sequence of length $N$. We thus estimate the local EP rate given by Eq.~\eqref{Eq:local_EP} by computing $\sigma_i = \tilde{\sigma}_i / \Delta t$, which would be valid for small enough $\Delta t$.

\begin{figure} [b!]
	\center
	\includegraphics[width=1\linewidth]{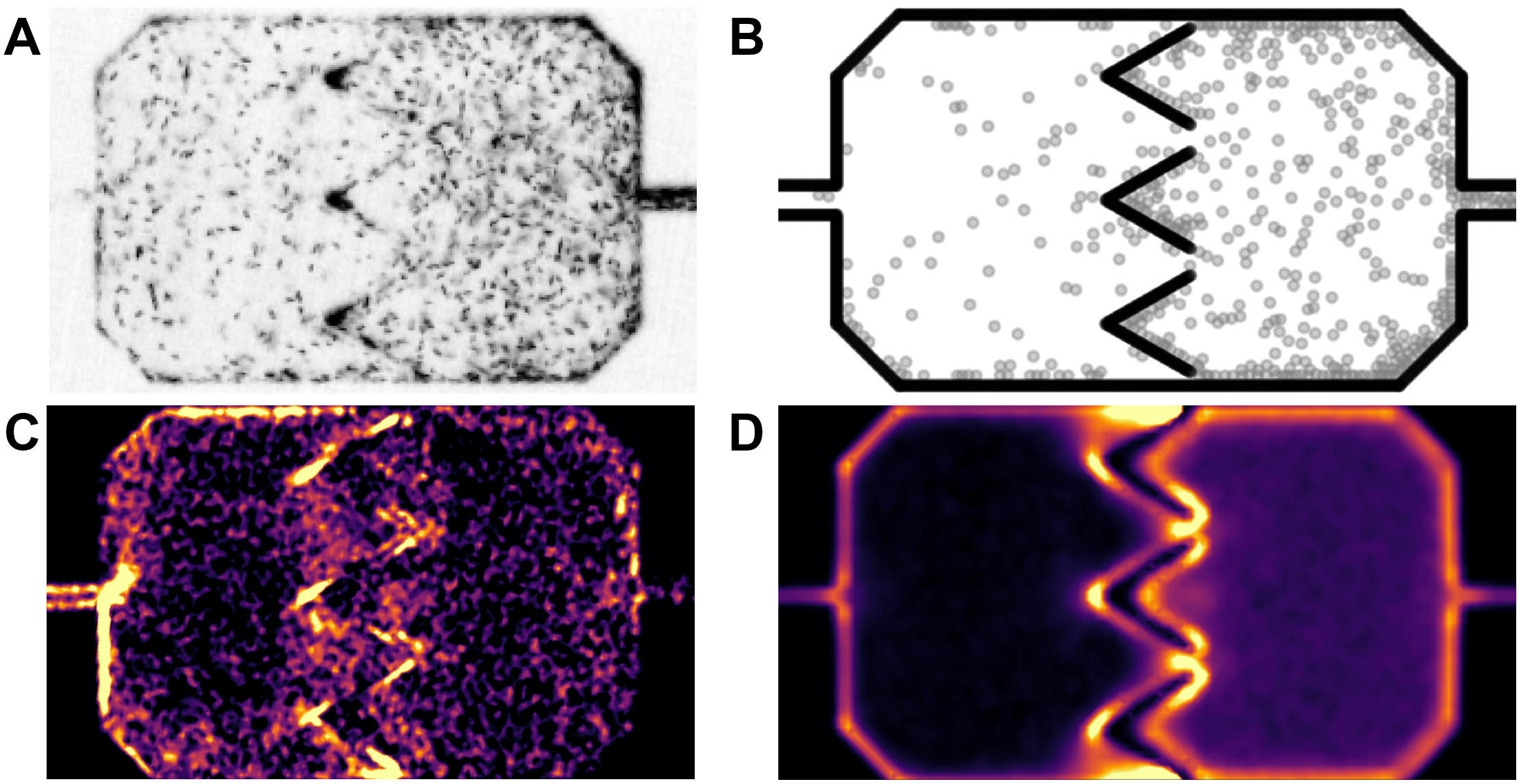}
	\caption{Local entropy production of E. Coli [snapshot in (A) and EP in (C)] and ABPs [snapshot in (B) and EP in (D)] in a rectifying cell. In the lower panels, the EP is captured with $2 \times 2$ blocks, with a brighter color indicating larger EP. The particle distributions show higher particle density on the right side of the cell driven by the funnels and particle accumulations along the walls.
	}
	\label{Fig:bacteria}
\end{figure}

\begin{figure*} [t!]
	\center
	\includegraphics[width=0.7\linewidth]{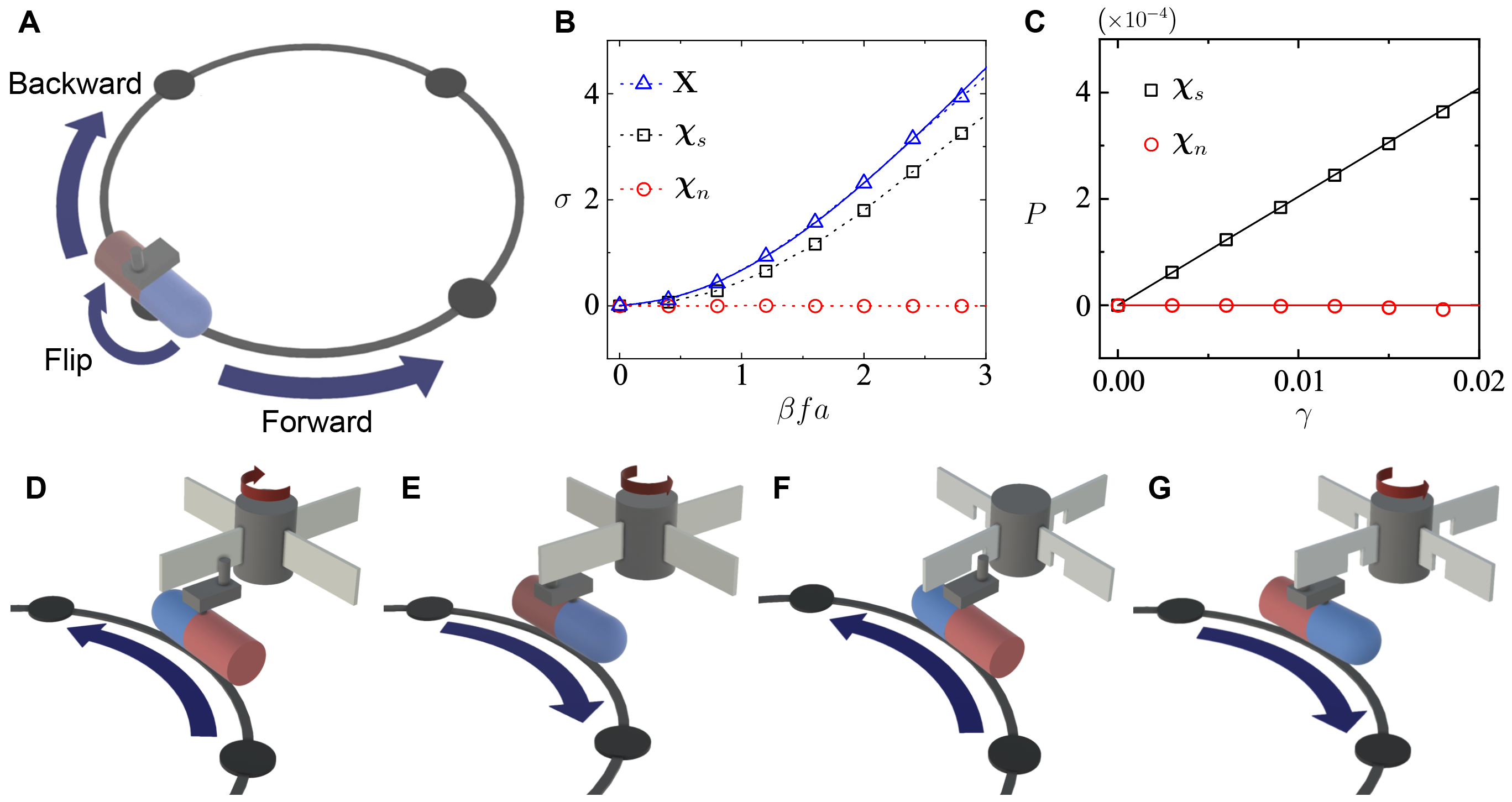}
	\caption{Entropy production and work extraction. (A) A run-and-tumble particle on a ring with four sites. The particle randomly flips its orientation, propels along its orientation, and experiences thermal noise. (B) EP measured with full information of the system ${\bf X}$ (blue triangles), occupancy and particle orientation on two sites $\boldsymbol\chi_s$ (black squares), and only occupancy on two sites $\boldsymbol\chi_n$ (red circles). The blue solid line is the global EP directly measured by accounting for heat dissipation: The entropy increases by $f a/T$ for forward motion, and decreases by the same amount for backward motion, where $f$ is the propulsion force, $a$ is the lattice constant, and $T$ is the temperature of the bath. No change in entropy is associated with the flipping of orientation. (C) The power $P \equiv \langle W \rangle / \tau$ extracted by the work extraction mechanisms shown in (D)-(G) when $\ln(w_f / w_b) = \beta f a = 0.25$, where $w_f$ and $w_b$ are the forward and backward rates and $\beta = (k_\mathrm{B} T)^{-1}$. Solid lines are exact results, see SI for a derivation. The mechanism that couples to $\boldsymbol\chi_n$, shown in (D) and (E), is unable to extract work, while the mechanism that couples to $\boldsymbol\chi_s$, shown in (F) and (G), has a finite positive slope in the small $\gamma$ limit. (D) and (E): interaction between the object and the mechanism coupling to $\chi_n$. (F) and (G): interaction between the object and the mechanism coupling to $\chi_s$.}
	\label{Fig:toy_model}
\end{figure*}

{\it Application to experiments}.---
Our procedure is applicable to physical and biological systems as well as to simulations. In Fig.~\ref{Fig:bacteria}, we show a snapshot of part of an experimental system in which E. coli are circulating in a network of chambers where funnels rectify the bacterial motion creating flows and density heterogeneities. We also show an analogous simulation with ABPs. Using the same protocol as in Fig.~\ref{Fig:ABP}, with 
$2 \times 2$ blocks, we see in the lower panels that the entropy production is largest at the funnel tips where the particle trajectories bifurcate, and at the boundary walls. There are also contributions from collisions in the left and right halves of the chamber with more entropy produced in the higher density regions, as expected.

{\it Work extraction}.---
Although the map of the local EP constructed with our measure provides ample information on how the system is driven out of equilibrium, its physical significance, in terms of its relationship to other physical observables, remains to be elucidated. We start by noting that the local EP is \emph{not} proportional to the heat dissipated by the local degrees of freedom~\cite{crooks_2019,fodor2021irreversibility,fodor2021active}. In what follows, we employ a simple model to argue that the local EP estimated by our measure is directly related to the work that can be extracted from a given region of the system.

To proceed, we consider a work extraction mechanism coupled to an active system and demonstrate that our measure can be used to identify which degrees of freedom must be considered (coupled to) in order to extract net work. Note that the extraction of work from active matter has received much attention recently~\cite{pietzonka2019autonomous,ekeh2020thermodynamic,speck2016stochastic,speck2018active} and our proposal is particularly relevant in the context of these related studies. Specifically, in this work, we consider a run-and-tumble particle (RTP) on a ring with four discrete sites with lattice spacing $a$, as shown in Fig.~\ref{Fig:toy_model}A, coupled to a thermal reservoir with constant temperature $T$. The particle flips randomly between two possible orientations, and jumps to the neighboring sites either driven by thermal noise or by a thrust $\mathbf{f}$ exerted in the direction of its orientation. We impose the local detailed balance condition to the particle motion, and the thrust makes forward motion more likely than backward motion, thus breaking TRS (see SI for details). For this model, we measure the EP for three different degrees of freedom: (i) the state of the entire system $X$, (ii) particle occupancy ${\chi}_n$ on two adjacent sites, and (iii) the particle occupancy and orientation ${\chi}_s$ on the same adjacent sites.

The results for the three measurements are shown in Fig.~\ref{Fig:toy_model}B. As expected, we observe dramatic dependence of the EP on the choice of degrees of freedom. The global EP, estimated by tracking the total state of the system $X$ (blue triangles), fully recovers the entropy production measured by accounting for the heat dissipation (solid blue line). In contrast, the local EP for $\chi_s$ (black squares) obtained by tracking the occupancy and orientation on two adjacent sites only partially captures the global EP, while the local EP for $\chi_n$ (red circles) obtained by tracking only the occupancy of two adjacent sites, but not the particle orientation, is zero, since the resulting trajectory is time-symmetric.

We now argue that the local EP is directly related to the amount of work that can be extracted by coupling to the tracked degree of freedom. That is, if a mechanism is weakly coupled to the degree of freedom with a small coupling strength $\gamma$, work can be extracted at the linear order in $\gamma$ only if a positive EP is obtained when tracking the degrees of freedom in the unperturbed state. To illustrate this, we consider an RTP with an asymmetric ``key'' fixed to its head. As shown in Fig.~\ref{Fig:toy_model}D-G, we introduce two turnstile-like mechanisms, one that couples to ${\chi}_n$, the other to ${\chi}_s$. Both mechanisms extract work when rotating counterclockwise, while releasing energy when rotating clockwise. The mechanism shown in Fig.~\ref{Fig:toy_model}D and E couples to $\chi_n$ and rotates whenever the particle moves between two sites where the turnstile arm is placed. By contrast, the mechanism depicted in Fig.~\ref{Fig:toy_model}F and G is orientation-specific - the turnstile rotates with the particle when it is oriented in clockwise-direction, but does not interact with the particle when the particle is oriented counterclockwise; it thus couples to $\chi_s$. The power harnessed by the mechanisms is shown in Fig.~\ref{Fig:toy_model}C, plotted with respect to the coupling strength $\gamma$, and clearly shows that the mechanism coupled to ${\chi}_n$ (red circles) does not harness work, while for the mechanism coupled to ${\chi}_s$ (black squares) the work extracted is linearly increasing with $\gamma$ and thus finite.

This observation can be generalized to any degrees of freedom coupled weakly to a work extraction mechanism based on nonequilibrium linear response theory~\cite{maes_response_2020}. Given that the system is weakly perturbed by the mechanism, the average power recorded by the work extraction mechanism during a time interval $t$ is given by $\langle P \rangle = \frac{\gamma}{2t} \langle S ( \boldsymbol\chi ) \tilde{W} ( \boldsymbol\chi ) \rangle_{\boldsymbol\chi} + {\cal O} (\gamma^2)$, where $S(\boldsymbol{\chi})$ and $\gamma \tilde{W}( \boldsymbol\chi )$ are the entropy produced and the work extracted by the mechanism, respectively, for a particular realization of $\boldsymbol\chi$~\cite{crooks_2019} (see SI for details). 
Note that $\tilde{W} (\boldsymbol\chi)$ is of order ${\cal O}(\gamma^0)$ in the small $\gamma$ regime, and the angled brackets indicate averaging over the steady-state measure for $\boldsymbol\chi$. Since entropy production is strictly zero for any time-symmetric trajectory $\boldsymbol\chi$, this equation indicates that the average extractable work is zero if the (local) EP measured for $\boldsymbol\chi$ is zero. 
\footnote{If $\boldsymbol{\chi}$ were Markovian, alternatively, one could have arrived at the same conclusion from the thermodynamic uncertainty relation, $\langle \tilde{W} (\boldsymbol{\chi})\rangle^2 \leq \mathrm{Var} [ \tilde{W}(\boldsymbol{\chi})] S(\boldsymbol{\chi})$~\cite{barato2015thermodynamic,horowitz2017proof}.}

To conclude, we have introduced a protocol to measure local EP by devising an effective information-theoretic measure, and demonstrated it on several numerical and experimental systems. We have shown that it is possible to identify which regions of the system are driven out of equilibrium in simulations and experiments, how these depend on the degrees of freedom being tracked and their relation to the locally extractable work. These results show that the local EP is an universal and powerful tool for studying nonequilibrium many-body systems, which extends recent efforts to apply tools of information theory to study physical systems~\cite{martiniani2019quantifying,avinery2019universal,martiniani2020correlation}. Note that the analysis of stochastic fluctuations by our approach, in experimental systems, may reveal the existence of nonequilibrium phenomena even when they are not easily distinguished from equilibrium processes. It would be interesting to apply our method to study various aspects of glassy dynamics, biological systems, and externally driven systems.

\emph{Acknowledgements} We thank Yariv Kafri for useful and insightful discussions, and Elsen Tjhung for clarifications regarding the field theoretic simulations. We thank Maria Jose Reyes for preparation of the figures. This work was supported by the National Science Foundation (NSF) Physics of Living Systems Grant No. 1504867. S.~M. acknowledges support from the Simons Foundation Faculty Fellowship and from the Simons Center for Computational Physical Chemistry at NYU. D.~L. thanks the U.S.-Israel Binational Science Foundation (Grant No. 2014713) and the Israel Science Foundation (Grant No. 1866/16). P.~M.~C. was supported partially by the Materials Research Science and Engineering Center (MRSEC) Program of the NSF under Award No. DMR-1420073 for data analysis. P.~M.~C. and B.~G. were partially supported by DOE SC-0020976 for numerical simulations. S.~R. and D.~L. were also supported by the National Research Foundation of Korea (2019R1A6A3A03033761). R.~H.~A and T.~.V.~P. were supported by the NSF through the Center for the Physics of Biological Function (PHY-1734030). A.~S. and S.~M. acknowledge the Minnesota Supercomputing Institute for compute time.

\bibliographystyle{apsrev4-1}
\bibliography{bibliography}

\end{document}


\title{Supplemental Information of Manuscript ``Play. Pause. Rewind. Measuring local entropy production and extractable work in active matter''}

    \author{Sunghan Ro}
	\thanks{Equal contribution.}
	\affiliation{Department of Physics, Technion-Israel Institute of Technology, Haifa 3200003, Israel}

	\author{Buming Guo}
	\thanks{Equal contribution.}
	\affiliation{Center for Soft Matter Research, Department of Physics, New York University, New York 10003, USA}
	
	\author{Aaron Shih}
	\affiliation{Courant Institute of Mathematical Sciences, New York University, New York 10003, USA}
	\affiliation{Department of Chemical Engineering and Materials Science, University of Minnesota, Minneapolis, Minnesota 55455, USA}
	
	\author{Trung V. Phan}
	\affiliation{Department of Physics, Princeton University, Princeton 08544, New Jersey, USA}
	
	\author{Robert H. Austin}
	\affiliation{Department of Physics, Princeton University, Princeton 08544, New Jersey, USA}
	
	\author{Dov Levine}
	\email{dovlevine19@gmail.com}
	\affiliation{Department of Physics, Technion-Israel Institute of Technology, Haifa 3200003, Israel}
	
	\author{Paul M. Chaikin}
	\email{chaikin@nyu.edu}
	\affiliation{Center for Soft Matter Research, Department of Physics, New York University, New York 10003, USA}
	
	\author{Stefano Martiniani}
	\email{sm7683@nyu.edu}
	\affiliation{Center for Soft Matter Research, Department of Physics, New York University, New York 10003, USA}
    \affiliation{Courant Institute of Mathematical Sciences, New York University, New York 10003, USA}
    \affiliation{Department of Chemical Engineering and Materials Science, University of Minnesota, Minneapolis, Minnesota 55455, USA}
    \affiliation{Simons Center for Computational Physical Chemistry, Department of Chemistry, New York University, New York 10003, USA}

\maketitle


\tableofcontents{}

\section{Information theoretic measure for entropy production rate}

The Kullback-Leibler (KL) divergence between the probability of observing a nonequilibrium process forward in time and the probability of observing its time-reversed realization is equal to the entropy production rate (EP) of that process~[5]. In other words, it is sufficient to evaluate the KL divergence between the ensembles of the forward and the time-reversed trajectories of a system to measure the EP of the system.

Here, we present our information-theoretic measure for estimating the EP of a stationary and ergodic Markov source with finite alphabet and finite memory. We will show numerically that our measure converges to the true EP of a simple Markov process within a much smaller number of samples than alternative measures of the same kind, namely those proposed by Ziv and Merhav~[28], and by Rold{\'a}n and Parrondo~[20]. 

To start, let us consider a random trajectory ${\bf X}_{1, N} = (X_1 X_2 \cdots X_N)$ where $X_i$ denotes certain degrees of freedom of the system at time $t=(i-1)\tau$, where $\tau$ is the sampling interval. We denote ${\cal P}[{\bf X}_{1, N}]$ the probability density function (PDF) of observing the trajectory ${\bf X}_{1,N}$, and ${\cal P}^R[{\bf X}_{1, N}]$ the PDF of observing its time-reversed realization ${\bf X}^R_{1, N} = (X_N X_{N-1} \cdots X_1)$. Without loss of generality, we only consider (for simplicity) variables that are even under the time-reversal operation. 
We can thus introduce our measure for EP
\begin{equation}\label{Eq:sigma}
	\sigma [{\bf X}] = \frac{\ln n}{n} \left[ {\cal C}({{\bf X}}_{1,n} || {\bf X}_{n+1,2n}^R) - {\cal C}({\bf X}_{1,n}||{\bf X}_{n+1,{2n}}) \right]~.
\end{equation}
Here, we are assuming that the total length of the sequence is $N = 2n$, and ${\cal C}({\bf Y}||{\bf Z})$ is the cross-parsing complexity obtained by Lempel-Ziv (LZ) coding of ${\bf Y}$ using ${\bf Z}$ as a dictionary~[28, 29]. As noted in the main text, $\sigma[{\bf X}] /\tau$ converges to the entropy production rate in the limit of $N \rightarrow \infty$ and $\tau \rightarrow 0$.

To describe the LZ cross-parsing in detail, we consider a pair of sequences ${\bf Y} = (012101110230)$ and ${\bf Z} = (301201110310)$. The cross-parsing is performed by sequentially drawing the longest consecutive array of symbols from the sample sequence ${\bf Y}$ which can also be found in the dictionary sequence ${\bf Z}$. In this example, the result of the parsing is $(012, 10, 1110, 2, 30)$ which gives a complexity ${\cal C}({\bf Y}|| {\bf Z}) = 5$. To derive Eq.~\eqref{Eq:sigma}, we adopt from Ref.~[28, 29] the following expressions that are valid in the asymptotic limit $n \to \infty$
\begin{equation} \label{eq:si_backward}
   \lim_{n \to \infty} \left( \frac{\ln n}{n} {\cal C}({{\bf X}}_{1,n} || {\bf X}_{n+1,2n}^R) - H[ {\bf X}_{1,n}] - \mathrm{KL}[{\cal P} || {\cal P}^R] \right) = 0~,
\end{equation}
and
\begin{equation} \label{eq:si_forward}
   \lim_{n \to \infty} \left( \frac{\ln n}{n} {\cal C}({{\bf X}}_{1,n} || {\bf X}_{n+1,2n}) - H[ {\bf X}_{1,n}] \right) = 0~.
\end{equation}
where $H[{\bf X}_{1,n}]$ is the entropy rate of the sequence satisfying
\begin{equation}
    \lim_{n \to \infty} \left( H[{\bf X}_{1,n}] + \frac{1}{n} \ln ( {\cal P}[{\bf X}_{1,n}] ) \right) = 0,
\end{equation}
and the KL divergence reads
\begin{equation}
    \mathrm{KL} [{\cal P} || {\cal P}^R] = \frac{1}{n} \sum_{{\bf X}_{1,n}} {\cal P}[{\bf X}_{1,n}] \ln \frac{ {\cal P} [{\bf X}_{1,n}] }{ {\cal P}^R[{\bf X}_{1,n}] }~.
\end{equation}
where the summation is performed over all possible realizations of ${\bf X}_{1,n}$. Subtracting Eq.~\eqref{eq:si_forward} from Eq.~\eqref{eq:si_backward} we arrive at our measure Eq.~\eqref{Eq:sigma}.

\begin{figure*}[t]
	\includegraphics[width=1.0\textwidth]{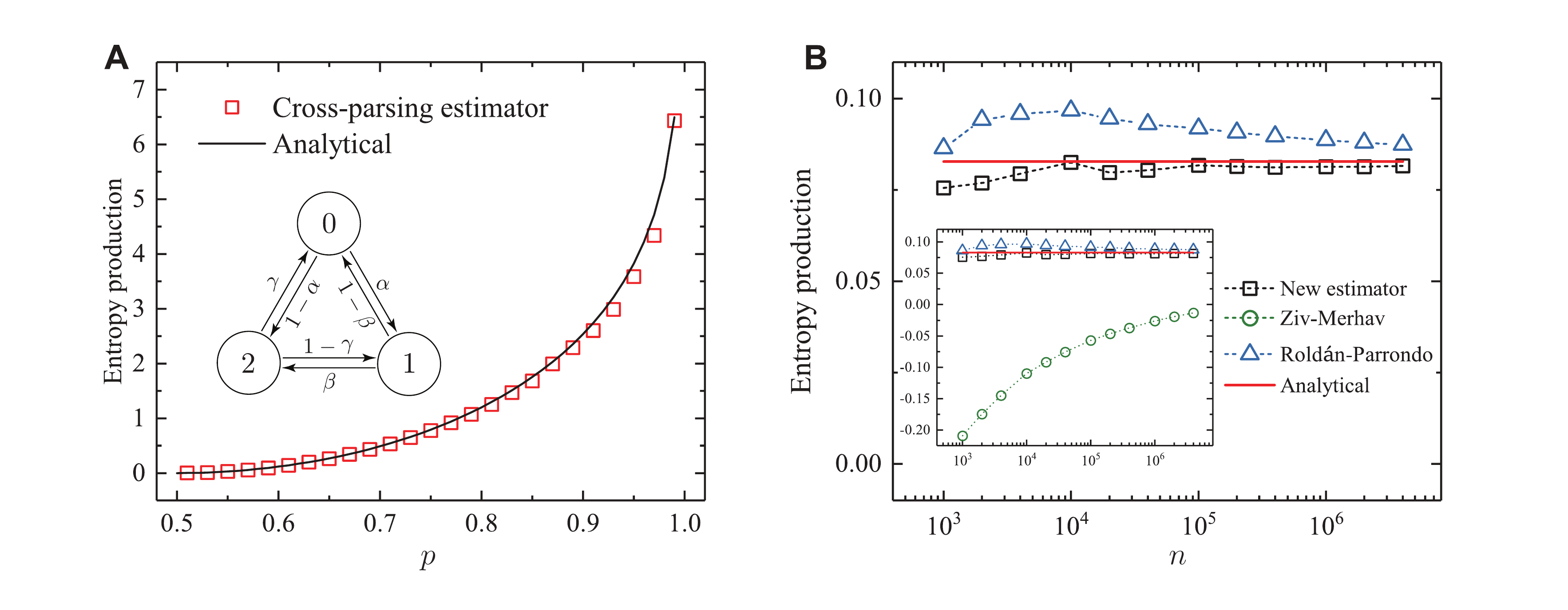} 
	\caption{(A) Entropy production of a three-state Markov model calculated using our new modified cross-parsing estimator (red squares) and its analytical curve given by Eq.~\ref{eq:si_markov_ep} (solid black). On the x-axis $p =  \alpha = \beta = \gamma$ is the transition probability of $0 \to 1 \to 2 \to 0$. Data shown are for a sequence of length $n = 10^6$. (B) EP with respect to the number of samples $n$ with comparisons between our new estimator (black squares), the original Ziv-Merhav (green circles) and the modified version by Rold\'an-Parrondo (blue triangles). Red line is the analytical result. Test performed using three-state Markov with $\alpha = 0.5$, $\beta = 0.7$, and $\gamma = 0.6$, which was adopted in Ref.~[20].
	}
	\label{FigS:Markov_test} 
\end{figure*}

To provide an informal justification of why our measure works, we present a simple example of a memoryless three-state Markov process, whose states are represented with symbols 0, 1, and 2. For each state, the transition probability $W(i \to i+1) = p$, and the transition probability for its reverse is given as $W(i+1 \to i) = 1-p$. We use a cyclic convention for the states so that state $i$ is the same as state $i+3$ (see Fig.~\ref{FigS:Markov_test}A, with $\alpha = \beta = \gamma = p$). If $p >1/2$, it is more likely to observe the cycle $0 \to 1 \to 2 \to 0$ than the cycle $0 \to 2 \to 1 \to 0$, and vice versa. Note that in the time-reversed sequence the transition $W^R(i \to i+1) = W(i+1 \to i) = 1-p$ and $W^R(i+1 \to i) = W(i \to i+1) = p$. 
Lastly, the entropy produced per step, $\sigma$, can be easily calculated by using the property of the memoryless Markov process as
\begin{align} 
    \nonumber \sigma =&\, \sum_j W(i \to j) \ln \frac{W(i \to j)}{W(j \to i)} \\
    =&\, p \ln \frac{p}{1-p} + (1-p) \ln \frac{1-p}{p}~. \label{eq:si_markov_ep}
\end{align}

Next, we evaluate our measure on a trajectory obtained from this system and compare it to Eq.~\eqref{eq:si_markov_ep}. Consider the trajectory ${\bf X}_{1, N} = (X_1 X_2 \cdots X_N)$ with $N = 2n$, where $X_i$ is the state of the $i$-th observation. According to the definition of our measure, we divide the sequence into two equal parts and compress the first half ${\bf X}_{1,n}$ with the forward and the backward realizations of the second half ${\bf X}_{n+1,2n}$. Suppose that the word ${\bf X}_{1,\ell}$ is drawn from the first half, corresponding to the first $\ell$ symbols of ${\bf X}_{1,n}$. The first symbol, $X_1$, can either be 0, 1, or 2 with equal probability, and in the rest of the word there will be on average $n_+ = p (\ell-1)$ of $i \to i+1$ transitions and $n_- = (1-p) (\ell-1)$ of $i \to i-1$ transitions. The probability $P(\ell)$ that a word drawn from a randomly chosen location of the second sequence $\mathbf{X}_{n+1, 2n}$ is the same as ${\bf X}_{1,\ell}$ will be
\begin{equation}
    P(\ell) \simeq \frac{1}{3} q^{p(\ell-1)}(1-q)^{(1-p)(\ell-1)}~,
\end{equation}
where $q = p$ when using a forward realization of the sequence ${\bf X}_{n+1, 2n}$ as the dictionary, and $q = 1-p$ when using its reverse, ${\bf X}_{n+1, 2n}^R$, as the dictionary. 
If the length of the word is large enough, we can approximate $P(\ell)$ as
\begin{equation}
    P(\ell) \simeq e^{-s(p,q) \ell}~,
\end{equation}
where the cross entropy $s(p,q) = - p \ln q - (1-p) \ln(1-q)$. 

If the sequence drawn from the second half does not match ${\bf X}_{1,\ell}$, one can look for a match at other locations in the sequence.  When doing this, the number of independent words $N_w(\ell)$ of length $\ell$ that can be drawn from the sequence is bounded by
\begin{equation} \label{Eq:si_Nw}
    \frac{n}{\ell} < N_w (\ell) < n - \ell~.
\end{equation}
Accordingly, the probability for not being able to find a match anywhere in the database satisfies
\begin{align}
    \nonumber P_\mathrm{NOT} (\ell) \simeq& \, \left[ 1 - P(\ell) \right]^{N_w(\ell)} \\
    \simeq& \, \exp \left[ - N_w(\ell) e^{-s(p,q) \ell} \right]. \label{eq:PNOT}
\end{align}
Note that $P_\mathrm{NOT}(\ell)$ is an increasing function of $\ell$, which converges to 1 for the sufficiently lengthy words. 

Next, to calculate the number of phrases obtained by the cross-parsing appearing in Eq.~\eqref{Eq:sigma}, we need to evaluate the average length of the longest matching words that are drawn at random locations from the first half and found anywhere in the second half. Here, it is crucial to realize that the probability $P_\mathrm{longest}(\ell)$ for the longest match to be of length $\ell$ satisfies 
\begin{equation} \label{eq:si_PL}
    P_\mathrm{longest}(\ell) = P_\mathrm{NOT} (\ell + 1) - P_\mathrm{NOT} (\ell).
\end{equation}
To understand this relation, it is intuitive to think about a set of sequences $S_{\ell+1}$ that have no match of length $\ell+1$ that can be found. From this set, if we subtract a subset $S_\ell$, consisting of the sequences for which a match of length $\ell$ cannot be found either, we are left with sequences for which we can find a match of length $\ell$ but not $\ell+1$. Thus, $\ell$ is the longest size of the match found from $S_{\ell+1} \setminus S_{\ell}$ (where $\setminus$ denotes the set theoretic difference), which is expressed as Eq.~\eqref{eq:si_PL} in terms of probabilities i.e., $P_\mathrm{longest}(\ell) = (|S_{\ell+1}| - |S_\ell|)/|S_{n}| = P_\mathrm{NOT} (\ell + 1) - P_\mathrm{NOT} (\ell)$. Using this relation, we evaluate the average length of the longest match as
\begin{align}
    \nonumber \langle \ell \rangle_\mathrm{longest} =& \, \sum_{\ell = 0}^{n-1} \ell P_\mathrm{longest} (\ell) \\
    \nonumber =& \, \sum_{\ell = 0}^{n-1} \left[ P_\mathrm{NOT}(n) - P_\mathrm{NOT}(\ell) \right] \\
    \simeq& \, \sum_{\ell=0}^{n-1} \left[ 1 - P_\mathrm{NOT}(\ell) \right] \label{eq:si_ell}
\end{align}
where the second equality can be found by expanding the sum, and we adopted an approximation $P_\mathrm{NOT}(n) \simeq 1$ for a sufficiently large $n$. 
Rewriting $P_\mathrm{NOT}(\ell)$ as $P_\mathrm{NOT}(\ell) \simeq \exp \left[ - e^{ \ln N_w(\ell) - s(p, q) \ell } \right]$ using Eq.~\eqref{eq:PNOT}, we can see that $\ell$ is a function rapidly varying around 
\begin{equation}
    \ell^\ast = \frac{\ln N_w(\ell)}{s(p,q)}
\end{equation}
and converges to zero as $\ell$ becomes smaller than $\ell^\ast$, and converges to one as it becomes larger than $\ell^\ast$. Based on this observation and Eq.~\eqref{eq:si_ell}, we approximate the average match length as $\langle \ell \rangle_\mathrm{longest} \simeq \ell^\ast$. 

Now we introduce a simplified notation for the number of phrases, written as ${\cal C}(p,q)$, denoting the result of the parsing by the forward sequence when $q=p$, and by the backward sequence when $q = 1-p$.
Using the relation ${\cal C} = n / \langle \ell \rangle_\mathrm{longest} \simeq n / \ell^\ast$, and Eq.~\eqref{Eq:si_Nw}, we write the bound for the number of phrases as
\begin{equation}
   \frac{n s(p,q)}{\ln n + { \ln(1 - \ell^\ast /n) } } < {\cal C} (p, q) <  \frac{n s(p,q)}{\ln n - \ln \ell^\ast }~.
\end{equation}
Since $\ell^\ast$ scales as $\ln n$, in the large $n$ limit ${\cal C} (p, q)$ satisfies
\begin{equation}
    \lim_{n \to \infty} \frac{\ln n}{n} {\cal C} (p, q) = s(p,q). \label{eq:cross-entropy}
\end{equation}
Plugging Eq.~\eqref{eq:cross-entropy} into our measure, Eq.~\eqref{Eq:sigma}, we obtain
\begin{align}
    \nonumber \lim_{n \to \infty} \sigma [ {\bf X} ] =&\, \lim_{n \to \infty} \frac{\ln n}{n} \left[ {\cal C}(p,1-p) - {\cal C}(p,p) \right] \\
    =&\, p \ln \frac{p}{1-p} + (1-p) \ln \frac{1-p}{p}
\end{align}
which is equal to the EP of the system, Eq.~\eqref{eq:si_markov_ep}, that we aim to obtain using our measure. In Fig.~\ref{FigS:Markov_test}A, we compare EP obtained with our cross-parsing estimator to the analytic result given in Eq.~\eqref{eq:si_markov_ep} for various value of $p$, which show excellent agreements. 

Compared to the original measure of EP proposed by Ziv and Merhav~[28] and a modified version of the measure suggested by Rold{\'a}n and Parrondo~[20], our measure shows much faster convergence to the actual value of EP due to its symmetric construction. We demonstrate this for a three-state Markov process as shown in Fig.~\ref{FigS:Markov_test}B.

\section{Simulation on active matter}

\subsection{Off-lattice simulation}

Our numerical results on active matter are obtained with active Brownian particles (ABPs) in two-dimensional on-lattice and off-lattice systems. For off-lattice simulation, we have adopted the model studied by Fily and Marchetti~\cite{fily2012athermal}, where the motions of each particle are governed by the following Langevin equations:
\begin{eqnarray}
    \partial_t {\bf r}_i(t) &=& v_0 {\bf \hat{e}(\theta_i)} + \mu \sum_{j \neq i} {\bf F}_{ij}\,,  \\
    \partial_t \theta_i (t) &=& \eta_{i} (t)\,,
\end{eqnarray}
where the particle index $i$ runs from 1 to $N$. Here, ${\bf r}_i$ is the particle position, ${\bf \hat{e}(\theta_i)}$ is the particle orientation with the angle $\theta_i \in [0,2 \pi)$, $v_0$ is a particle propulsion speed, $\mu$ is a mobility, ${\bf F}_{ij}$ is a force exerted on the particle $i$ by the particle $j$, $\eta_i(t)$ is a rotational thermal noise following the Gaussian distribution with the zero mean and correlations of $\langle \eta_i (t) \eta_j (t') \rangle = 2 D_r \delta_{ij} \delta (t - t')$ with the rotational diffusion coefficient $D_r$. The simulation is performed with HOOMD-blue \cite{andreson2020hoomd} for GPU acceleration.
For interaction between the particles, we assign the following interaction forces
\begin{equation}
    {\bf F}_{ij} =  \frac{\epsilon}{2a} \left( 1 - \frac{r_{ij}}{2a} \right) \Theta \left(1 - \frac{r_{ij}}{2a}\right) \hat{\bf r}_{ij}
\end{equation}
with $r_{ij} \equiv | {\bf r}_j -  {\bf r}_i|$, $\hat{{\bf r}}_{ij} \equiv {\bf r}_{ij} / r_{ij}$, $\Theta(r)$ is a Heaviside step function, $\epsilon = 1$, and the particle radius $a=1$. 
For simulations on the motility induced phase separation (MIPS), we set the parameters as $N=4096$, $v_0=0.1$, $\mu = 1$, $D_r= 5 \times 10^{-4}$, the volume fraction is $\phi=0.4$, and the system size is controlled accordingly. The time step used in our simulation is 0.01, and we record the state of the system once every 1000 steps to construct the trajectories. This means that the particles travel around distance 1 at most between two consecutive recordings. Then we discretize the configurations by superimposing  a $256 \times 256$ square lattice on the simulation box, and designate each site (plaquette) as empty or occupied according to whether there is any of the centers of the particles in the site or not. 
With this choice, there is at most only one particle in each $2 \times 2$ block, and each particle moves at most a distance equivalent to one lattice side between two successive measurements. 

To measure the local entropy production around the droplet formed by MIPS, we fixed the center of mass of the particles at the center of our simulation box every frame so that the droplet does not diffuse around over the observation time. 
We then performed the aforementioned spatial discretization, and assigned each site an index 1 if it is occupied by a particle and 0 if it is empty. Then we construct the local state associated with each grid point by recording the occupancy state of $n \times n$ blocks where we chose $n = 2, 3$, and $4$ and measure the EP by plugging in the obtained trajectory into our measure Eq.~\eqref{Eq:sigma}. Finally, we obtain the EP of each grid point by averaging EPs measured from the blocks. For $2 \times 2$ blocks, we average out EPs obtained from all blocks containing the chosen grid point. For $3 \times 3$ block, we use the EP measured by the block whose center is located at the chosen grid point. For $4 \times 4$ blocks, we take an average over four blocks whose $2 \times 2$ center block contains the chosen grid point.

\begin{figure*}[t]
	\includegraphics[width=0.8\textwidth]{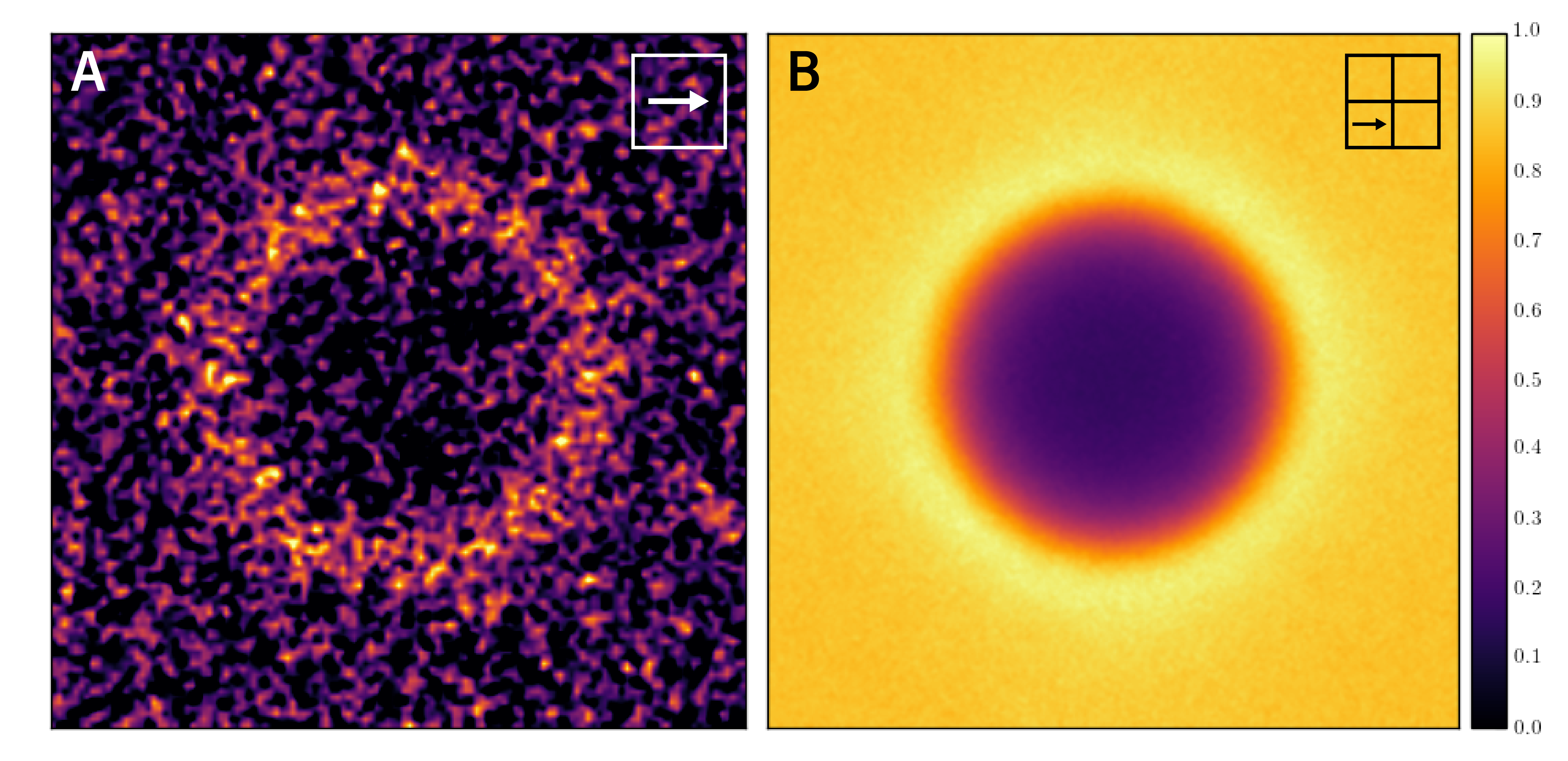} 
	\caption{Normalized local entropy production rate of active system undergoing MIPS, obtained by tracking orientation degrees of freedom. (A) EP with single block orientations. (B) EP with orientations in $2 \times 2$ block that captures the entropy production due to active propulsion in the dilute phase. Normalization factors are (A) $2.3 \times 10^{-3}$, and (B) $2.6 \times 10^{-1}$.
	} 
	\label{FigS:orien} 
\end{figure*}

Next, we present additional results obtained with the off-lattice simulation. 
In the main text, we reported the dependence of the EP on the degrees of freedom tracked, by comparing results with 
different local state representations (with occupancies in $n \times n$ blocks where $n= 2, 3$, and 4). 
We further support this observation with the map of EP obtained by measuring degrees of freedom including orientations of active particles. To start, as shown in the inset of Fig.~\ref{FigS:orien}A, we show an EP map obtained by constructing the local state using a particle occupancy and orientation at a single lattice site. For this, we represent the state of the lattice as one of the following five states: the particle is pointing upward, downward, left, right, or the site is empty. The map of EP obtained as a result presents relatively large value of EP at the boundary between the phases. This can be understood by considering that, at the boundary, the particle would remain at the same position while pointing in the direction of the dense phase, but would quickly leave the location if pointing toward other directions. In Fig.~\ref{FigS:orien}B, we show the EP obtained by considering the orientation state at $2 \times 2$ sites. In contrast to Fig.~\ref{FigS:orien}A, significantly larger EP is captured from the dilute phase. To understand this, we remind the reader that the particles are more likely to propagate forward than backward due to self-propulsion, thus breaking the time reversal symmetry, which is now captured by tracking the particle occupancy and orientation in a $2 \times 2$ block. This should be compared to the EP obtained by tracking only occupancy states in $2 \times 2$ blocks shown in Fig.~1B of the main text, which was not able to capture the EP in the dilute phase.

\begin{figure*}[t]
	\includegraphics[width=0.8\textwidth]{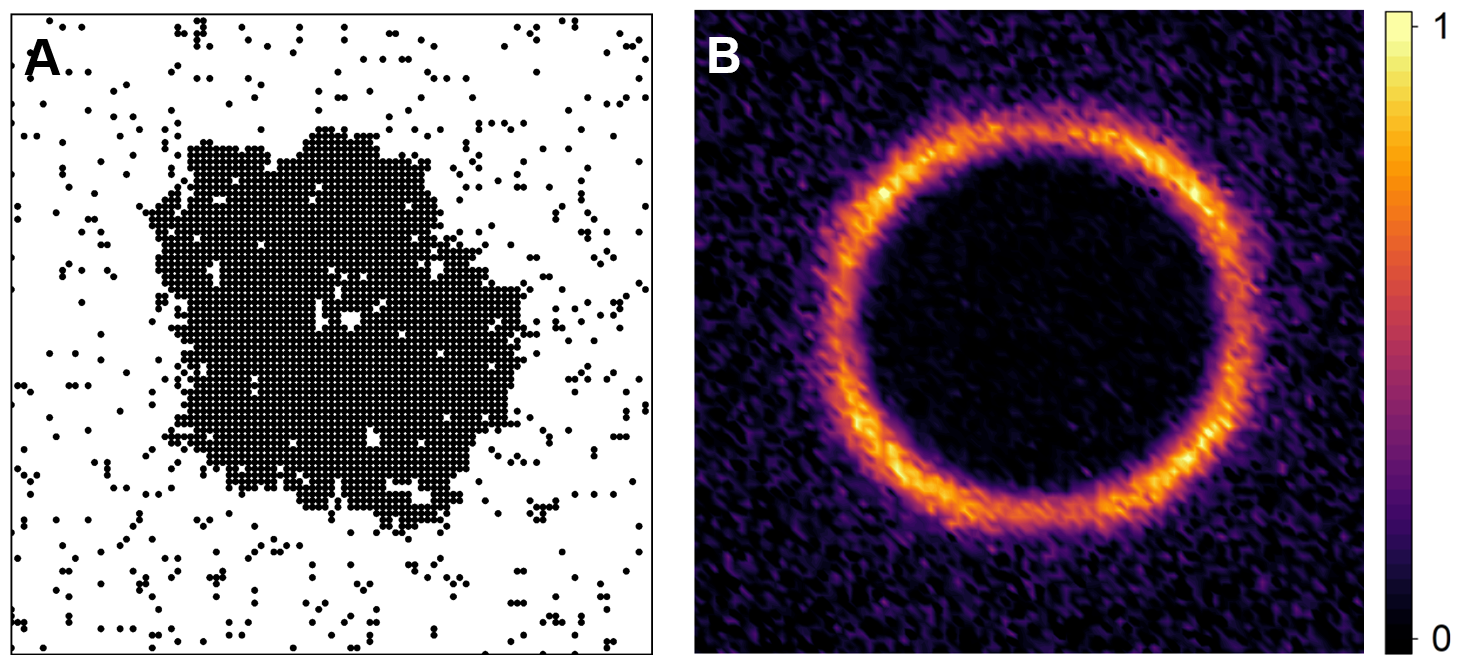} 
	\caption{(A) Simulation snapshot of on-lattice Active Brownian Particles forming a MIPS cluster and (B) entropy production rate measured from the lattice model. Normalization factor for the scale in (B) is $2.4 \times 10^{-2}$.
	} 
	\label{FigS:lattice} 
\end{figure*}

\subsection{On-lattice simulation}
We performed lattice simulations for active Brownian particles with excluded volume interaction on a two-dimensional $L \times L$ lattice. In these simulations, an orientation $\hat{e}(\theta) = (\cos \theta, \sin \theta)$ with $\theta \in [0, 2 \pi)$ is assigned to each particle, which randomly diffuses to $\theta + \eta$ with a rate $\alpha$ where $\eta$ is a random number drawn from a uniform distribution in $[- \pi/2, \pi/2]$. Due to active propulsion, each particle jumps from its original position $\vec{i}$ to one of its nearest neighbors $\vec{j}$ with a rate $W_{\vec{i}, \vec{j}} = \max[v \hat{e}(\theta) \cdot \vec{u}, 0]$ where $\vec{u} \equiv \vec{j} - \vec{i}$ is the displacement vector. Interaction between the particles prevents occupation of each site by more than one particle and thus hopping to a site occupied by another particle is prevented. If $v_0 / \alpha$ is large and the particle density is high enough, the particles induce phase separation into a dense and a dilute phases. 

In Fig.~\ref{FigS:lattice}A and \ref{FigS:lattice}B, we present a snapshot of the on-lattice simulation presenting MIPS and EP obtained from it. We set the parameters as $v_0 = 50$, $\alpha = 1.0$, $L = 100$, $\phi = 0.3$, with the number of simulation steps $8 \times 10^4$ where $\Delta t \equiv (\alpha + \sqrt{2} v_0)^{-1}$ is the unit time interval. Figure.~\ref{FigS:lattice}A presents a snapshot of the MIPS droplet obtained with these parameters. To obtain the map of EP shown in Fig.~\ref{FigS:lattice}B, similarly to the off-lattice case, we placed the center of mass of the particles at the center of the system, and recorded occupancy states in each $2 \times 2$ lattice block with the observation time interval set to $\Delta t$. The obtained time sequence is then plugged into our measure Eq.~\eqref{Eq:sigma} and the EP measured is then plotted at the center of the respective $2 \times 2$ block. The resulting map of EP is consistent with the result obtained with the off-lattice simulation, which supports the wide applicability of our measure.

\section{Active Model B}
Active Model B is a field-theoretical model of phase separation of a system without time-reversal symmetry  suitable for the description of conservative particles in nonequilibrium with scalar order parameter~[23].
The order parameter $\phi$ represents a reference phase of high density when $\phi=1$ and a reference phase of low density when $\phi=-1$. The dynamics of $\phi(\mathbf{r},t)$ obey:
\begin{align}
    \dot{\phi} = -\nabla\cdot\mathbf{J},\hspace{10mm}\mathbf{J}= -\nabla\mu+\mathbf{\Lambda}
    \label{AMB:current}
\end{align}
where $\mathbf{J}$ is the current and $\mathbf{\Lambda}$ is a Gaussian white noise field such that:
\begin{align}
    \mathbf{\Lambda}=\sqrt{2D}\mathbf{\Gamma}; \hspace{4mm}\langle{\Gamma_\alpha(\mathbf{r},t)\Gamma_\beta(\mathbf{r}',t')}\rangle=\delta_{\alpha\beta}\delta(\mathbf{r}-\mathbf{r}')\delta(t-t')
    \label{AMB:noise}
\end{align}
The chemical potential takes on the form:
\begin{align}
    \mu = \underbrace{a\phi + b\phi^3 - K\nabla^2\phi}_{\text{Model B}} + \underbrace{\lambda|\nabla\phi|^2}_{\text{Active term}}
    \label{AMB:mu}
\end{align}
When $\lambda=0$, this is simply Model B. However, the term $\lambda|\nabla\phi|^2$ is introduced to break time-reversal symmetry.

\begin{figure*}[t]
	\includegraphics[width=0.8\textwidth]{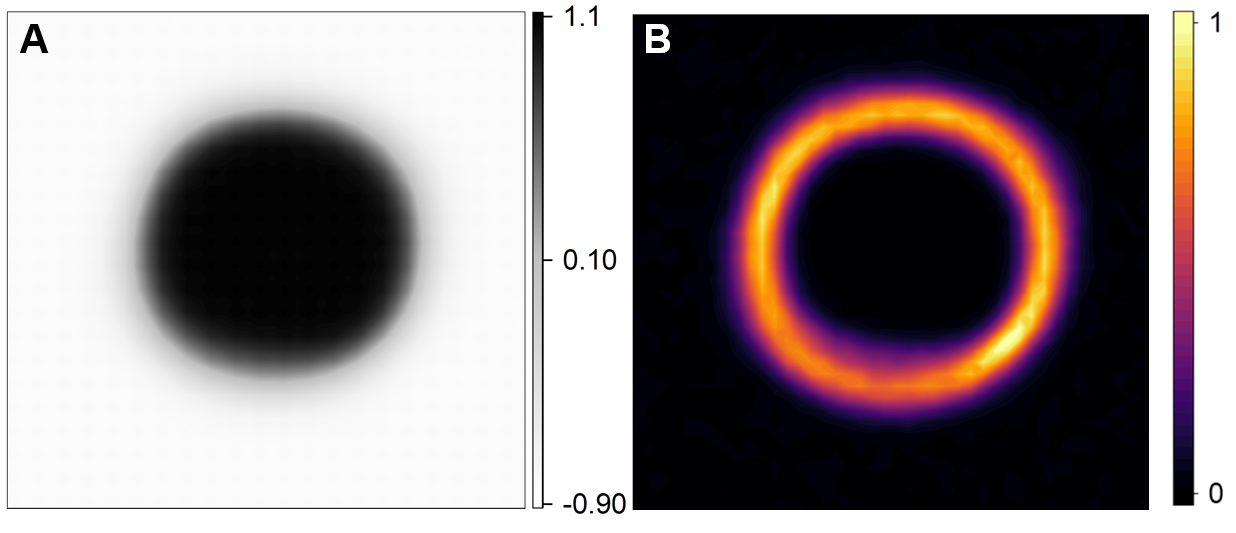} 
	\caption{(A) Density field featuring MIPS obtained by simulating Active Model B and (B) the respective entropy production rate. Parameters: $a = 1$, $b = -1$, $K = 1$, $\lambda = 1$, $D = 0.1$, and a unit time of $\mathrm{d}t = 10^{-3}$. Normalization factor for the scale in (B) is $1.3 \times 10^{-1}$.
	} 
	\label{FigS:AMB} 
\end{figure*}

In Fig.~\ref{FigS:AMB}, we present the density field obtained by simulating active model B and the corresponding entropy production rate. In simulating the equations for active model B, the gradient and divergence terms in Eqs.~\eqref{AMB:current} \&~\eqref{AMB:mu} are computed using a two-point stencil, e.g.:
\begin{align}
    (\nabla\phi)_i = \frac{\phi_{i+1}-\phi_{i-1}}{2\Delta}
    \label{AMB:grad}
\end{align}
where $\Delta = \Delta x, \Delta y$ as appropriate. The Laplacian $\nabla^2\phi$ in Eq.~\eqref{AMB:mu} is computed using an isotropic Laplacian kernel:

\begin{align}
    \nabla^2 = \frac{1}{\Delta x\Delta y}\left[\begin{matrix}-\frac{1}{2} & 2 & -\frac{1}{2}\\ 2 & -6 & 2\\-\frac{1}{2} & 2 & -\frac{1}{2}\end{matrix}\right]
    \label{AMB:laplacian}
\end{align}

Although the deterministic part of Eq.~\eqref{AMB:current} can be simplified into $\dot{\phi}=\nabla^2\mu$, in order to maintain detailed balance, we must compute the gradient and divergence separately, applying the scheme in Eq.~\eqref{AMB:grad} twice \cite{Banerjee_2017}.

To compute the entropy production rate, we adopt the convention of Ref.~[23] and estimate the local rate of entropy production $\mathcal{S}$ defined as
\begin{align}
    \mathcal{S} &= \lim_{\tau\to\infty}\frac{1}{\tau}\int_0^\tau dt\frac{\partial\phi}{\partial t}\circ\mu\\
    &= \frac{1}{\tau}\sum_t\Delta t\frac{\phi_{t+\Delta t}-\phi_{t-\Delta t}}{2\Delta t}\mu_t
    \label{AMB:entropy}
\end{align}

This is a Stratonovich time integral in which $\frac{\partial\phi}{\partial t}$ is discretized as in Eq.~\eqref{AMB:entropy} to be symmetric in time, such that the time integral remains unchanged if integrated over reverse time $t' = \tau-t$.
\section{A Brownian particle driven on a ring}

\begin{figure}[t!]
	\center
	\includegraphics[width=0.7\linewidth]{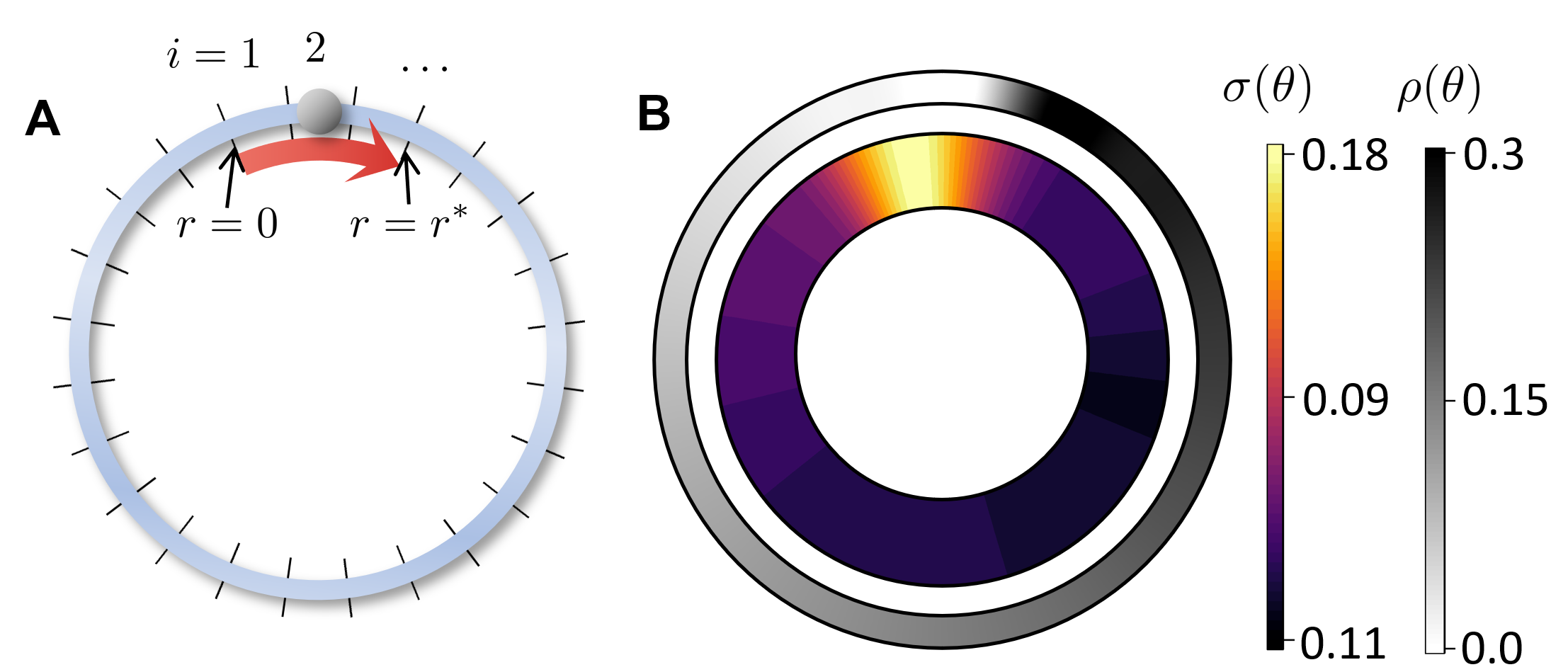}
	\caption{{\bf (A)} A Brownian particle on a ring driven by an external field applied to the upper end of the system, indicated by the red arrow. The ring is discretized into $L$ segments, with occupancy $0$ or $1$ in each segment. {\bf (B)} EP, $\sigma(\theta)$, is plotted as the inner ring and the probability density, $\rho(\theta)$, as the outer ring.}
	\label{Fig:scheme}
\end{figure}

To make a contrast between the global and local entropy production rates, we consider an elementary model: a Brownian particle on a circular track driven by a non-conserving force field depicted in Fig.~\ref{Fig:scheme}A. To describe the trajectory of the system state, we start by discretizing the domain by placing a grid with $L$ sites along the track. The state is recorded $N$ times with $\Delta t$ the time interval between measurements. The configuration $X(n)$ of the system at a given time $t = n \Delta t$ is a one dimensional array encoding the occupancy of $L$ sites. As there is only one particle, the sequence will have a 1 in correspondence of the occupied site, and 0s everywhere else. We denote a trajectory of configurations as ${\bf X} = X(0), X(\Delta t), \dots, X(n\Delta t)$.

Following the procedure suggested in the main text, we compare the recorded sequence ${\bf X}$ 
with the backward trajectory ${\bf X}^\mathrm{R}$. In the absence of an external driving, the motion of the Brownian particle is purely diffusive and the trajectories are statistically identical to each other. 
However, in the presence of an external driving exerted to the particle over a portion of the ring, the particle motion is biased in the direction of the driving, and the system exhibits TRSB. As a result, ${\bf X}$ can be distinguished from ${\bf X}^\mathrm{R}$. We can quantify the degree of TRSB by computing the EP rate as in Eq.~(2) of the main text.

The global entropy production rate can be accessed by accounting for the probability distribution and its current as suggested in Ref.~\cite{seifert2005entropy}. The resulting global EP is a single positive number whose non-zero value signifies that the system is out of equilibrium. This, however, obscures the system's spatial inhomogeneity; for example, we might expect TRS to be violated more strongly near where the force field is applied. To observe this, we must focus on the local dynamics of the system. Thus, we tile our $L$ grid with pairs of sites, and consider their occupancy states that can take on three possible values: 0 (unoccupied), and 1 or 2 depending on which site is occupied. We denote this (local) state by $\chi_i(t)$, where $i \in [1,L]$ specifies the grid point under observation. We then measure the local EP, $\sigma [\boldsymbol\chi_i]$, using our measure, the result of which is shown in Fig.~\ref{Fig:scheme}B. This spatial map highlights which region of the system is driven by the external field.

To support the validity of our simulation, we present an analytic prediction of the PDF and compare it to the results of the simulation. In the presence of the thermal bath with the temperature $T$ and driving force exerted on the particle, its equation of motion is given as
\begin{equation} \label{Eq:Ring_Eq_motion}
	\dot{r}(t) = \mu F(r) + \eta (t)~,
\end{equation}
where $r \in [0,2\pi R]$ is the position of the particle along the circumference of the ring with the radius $R$, $\eta(t)$ is a Gaussian random noise satisfying $\langle \eta (t) \rangle = 0$ and $\langle \eta (t) \eta (t') \rangle = 2D \delta (t - t')$ where the angled brackets denote the average over the noise realizations. The diffusion constant satisfies $D =  \mu \beta^{-1}$ with $\beta = (k_B T)^{-1}$ according to the fluctuation-dissipation theorem. We prescribe the force driving the particle as 
\begin{equation}
	F(r) = \begin{cases}
		F_0 & \mathrm{if} ~ 0 < r < r^\ast \\
		0 & \mathrm{if} ~ \mathrm{otherwise}
	\end{cases}.
\end{equation}
Here, $F_0$ is the driving strength and $r^\ast$ is the length of the driven regime. Note that this force cannot be obtained by taking a gradient of a potential field. In the main text, we have presented the EP and the probability density field (PDF) of the particle position obtained with dimensionless parameters of $2 \pi R / r^\ast = 12$, $\beta F_0 r^\ast = 5 \pi /2$, while dividing the circumference into 24 segments to evaluate EP.

\begin{figure}[t!]
	\includegraphics[width=0.4\textwidth]{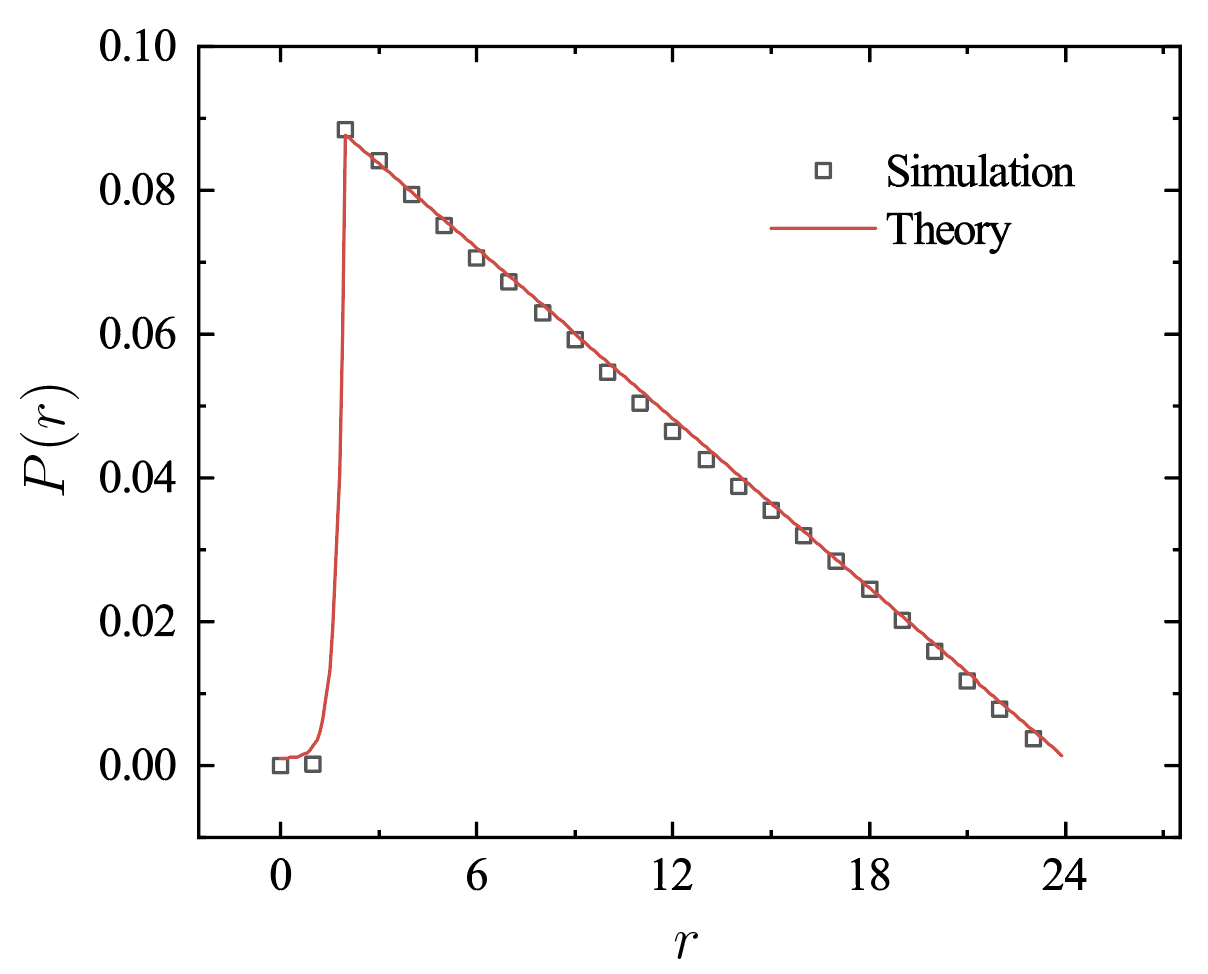} 
	\caption{Comparison between the probability density functions of a driven Brownian particle on a ring. The density obtained from simulation is presented in black symbols and the theoretical prediction given by Eqs.~\eqref{Eq:SI_ring_pdf1} and \eqref{Eq:SI_ring_pdf2} is drawn in a red solid line.} 
	\label{FigS:ring} 
\end{figure}

To evaluate the PDF in the stationary state, we consider a Fokker-Planck equation corresponding to the equation of motion Eq.~\eqref{Eq:Ring_Eq_motion} as
\begin{equation}
	\frac{\partial}{\partial t} P(r, t) = - \frac{\partial}{\partial r} [\mu F(r) P(r, t)] + D \frac{\partial^2}{\partial r^2} P(r, t)~.
\end{equation}

We solve the Fokker-Planck equation in the stationary state. First, in the driven region, the equation becomes
\begin{equation}
	0 = -\frac{\partial}{\partial r} \left[ \mu F_0 P(r) - D \frac{\partial}{\partial r} P(r) \right]~.
\end{equation}
Integrating this equation once, we obtain $J_0 = \mu F_0 P(r) - D \frac{\partial}{\partial r} P(r)$ where $J_0$ is the integration constant which is equivalent to the probability current according to the continuity equation. By integrating the expression obtained once more, we obtain
\begin{equation} \label{Eq:SI_ring_pdf1}
	P(r) = A e^{ \beta {F_0} r} - \frac{J_0}{D} \int_0^r \mathrm{d} r' ~ e^{\beta {F_0} (r - r')},
\end{equation}
where $A$ should be determined from the initial condition. This equation shows that in the interval $[0, r^\ast]$, the PDF changes exponentially with respect to $r$. 

Next, in the regime without driving, the Fokker-Planck equation becomes $J_0 = - D \frac{\partial}{\partial r} P(r)$, with the solution given as
\begin{equation} \label{Eq:SI_ring_pdf2}
	P(r) = B - \frac{J_0}{D} r~.
\end{equation}
We have a solution with three undetermined constants, $A, B$, and $J_0$, and three conditions imposed on the system: $\int_0^{2 \pi R} \mathrm{d} r' ~ P(r) = 1$, and the two solutions we got above should match with each other at $r = 0$ and $r = r^\ast$. This let us evaluate all the undetermined constants. Note that this solution indeed matches with the density field presented Fig.~\ref{FigS:ring}.
Note that we used the discrete grid points implemented to measure EP for evaluating the PDF, and there are respective numerical error if the PDF changes drastically within the grid size.

\section{Toy model and work extraction}

\subsection{Toy model}

\begin{figure}[b!]
	\includegraphics[width=0.5\textwidth]{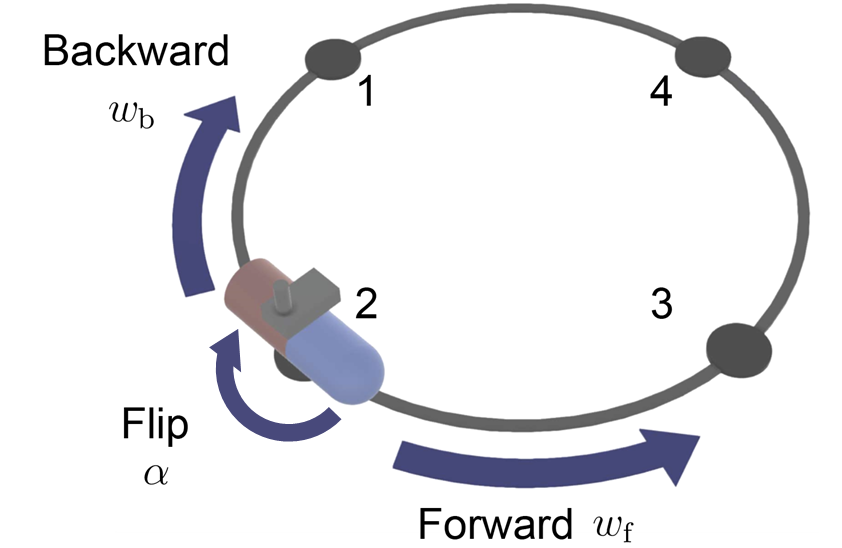} 
	\caption{Schematic depiction of the toy model and its dynamical transition rates.} 
	\label{FigS:toy_model} 
\end{figure}

Our toy model consists of a run-and-tumble particle and a ring with four sites $i = \{1, 2, 3, 4\}$ (see Fig.~\ref{FigS:toy_model}). The particle has an orientation along which it exerts a propulsion force $f$. The state of the whole system is specified by an array of four numbers $X = (s_1, s_2, s_3, s_4)$ where $s_i = +1(-1)$ if site $i$ is occupied by a clockwise-moving (counterclockwise-moving) particle, or $s_i=0$ if the site is empty. The particle flips its orientation with a rate $\alpha$, and jumps forward or backward with rates $w_f$ and $w_b$, respectively. Here, we impose the local detailed balance condition with a thermal bath with temperature $T$, which leads to $w_\mathrm{f} / w_\mathrm{b} = \exp(\beta fa)$ where $\beta \equiv (k_B T)^{-1}$ and $a$ is the spacing between the neighboring sites. For simplicity, we take $\alpha$ as the unit rate of our dynamics and set $w_\mathrm{f} / \alpha = \exp ( \beta f a /2)$ and $w_\mathrm{b} / \alpha = \exp ( -\beta f a /2)$.

As a result of the particle dynamics, the entropy of the bath changes. The entropy increases by $fa/T$ for forward motion, and decreases by the same amount for backward motion. No change in entropy is associated with the flipping. Therefore, the total change of entropy during a time interval can be measured by counting the difference between the forward and the backward motion within the interval. In Fig.~4B of the main text, we compare EP obtained by directly counting the number of forward and backward motions (blue line) to the EP calculated by applying our method on the trajectory ${\bf X}$ obtained by sampling $X$ once every time interval of $(\alpha + w_\mathrm{f} + w_\mathrm{b})^{-1}$ (blue symbols). As shown in the figure, the two results are in excellent agreement. 

In Fig.~4B and 4C of the main text, we examine the  dependence of EP on the degrees of freedom tracked by measuring only a fraction of the full state of the system X. In i) we include the occupancy and orientation on two adjacent sites $\chi_s \equiv (s_i, s_{i+1})$, and in ii) only the occupancy on these sites $\chi_n \equiv (n_i, n_{i+1})$ where $n_i \equiv |s_i|$. As depicted in the main text, EP measured with both the occupancy and orientation trajectory $\boldsymbol{\chi}_s$ is finite, and thus partially captures the time-reversal symmetry breaking in the system, but EP measured only with the occupancy trajectory $\boldsymbol{\chi}_n$ is zero since the forward and backward trajectories are indistinguishable. 

\subsection{Work extraction}

To further study the relationship between the EP associated with a particular set of degrees of freedom and the ability to extract work from those degrees of freedom, we consider an ensemble of trajectories of an observable $\chi$ obtained from a system of interest and describe a mechanism to extract work from it. Denoting the probability density function (PDF) for a trajectory ${\boldsymbol \chi}$ to be observed as ${\cal P}[{\boldsymbol \chi}]$,
we define the path-wise entropy production for a given 
trajectory ${\boldsymbol \chi}$ as
\begin{equation} \label{S}
	S[{\boldsymbol \chi}] = \ln \frac{{\cal P} [{\boldsymbol \chi}]}{{\cal P}^R [{\boldsymbol \chi}]}~.
\end{equation}
Here, ${\cal P}^R [{\boldsymbol \chi}] \equiv {\cal P}[\boldsymbol{\chi}^R]$ is the PDF for the time-reversed realization of ${\boldsymbol \chi}$. The average entropy production is written in the form of a Kullback-Leibler divergence as 
\begin{eqnarray}
	\nonumber \langle S \rangle &=& \int {\cal D} {\boldsymbol \chi} ~ {\cal P} [{\boldsymbol \chi}] \ln \frac{{\cal P} [{\boldsymbol \chi}]}{{\cal P}^R [{\boldsymbol \chi}]} \\
	\label{KL} &=& \left\langle \ln \frac{{\cal P} [{\boldsymbol \chi}]}{{\cal P}^R [{\boldsymbol \chi}]} \right\rangle~.
\end{eqnarray}
where the average is taken over noise realizations. We will show that the entropy production is directly related to whether one can extract work by coupling a mechanism to the degrees of freedom tracked by ${\boldsymbol \chi}$ or not.

Consider introducing a mechanism for extracting work from the system, where the coupling strength between the observable and the mechanism is $\gamma$. The work extracted by this mechanism from the degrees of freedom tracked by the trajectory ${\boldsymbol \chi}$ is given as
\begin{eqnarray} \label{W}
	W[{\boldsymbol \chi}] = \int_{\chi_i}^{\chi_f} \mathrm{d} \chi \cdot \mathbf{F} (\chi)~,
\end{eqnarray}
where we denote the state of the degrees of freedom at the initial and the final states as $\chi_i \equiv \chi(t_i)$ and $\chi_f \equiv \chi(t_f)$, respectively. 
In the end, we are interested in the average work, which is evaluated as 
\begin{equation} \label{Woriginal}
	\langle W \rangle = \int {\cal D} {\boldsymbol \chi} ~ {\cal P}[\gamma, {\boldsymbol \chi}] \int_{\chi_i}^{\chi_f} \mathrm{d} \chi \cdot \mathbf{F} (\chi)~.
\end{equation}
Here, we emphasized the fact that the PDF is, in general, a function of the coupling strength $\gamma$ between the mechanism and the system. Assuming that the PDF is an analytic function of $\gamma$, we can expand it about $\gamma$ to obtain
\begin{equation} \label{Pgamma}
	{\cal P} [\gamma, {\boldsymbol \chi}] = {\cal P} [{\boldsymbol \chi}] + \gamma {\cal P}' [\gamma, {\boldsymbol \chi}]|_{\gamma=0} + {\cal O}(\gamma^2)~,
\end{equation}
where ${\cal P} [{\boldsymbol \chi}]$ is the PDF observed without coupling the system to the work extraction mechanism. Inserting Eq.~\eqref{Pgamma} into Eq.~\eqref{Woriginal} and using the fact that $\mathbf{F} (\chi)$ is proportional to $\gamma$ in the weak coupling regime, we can expand the equation for the average work to obtain
\begin{equation} \label{Wlinear}
	\langle W \rangle = \int {\cal D} {\boldsymbol \chi}\, {\cal P}[{\boldsymbol \chi}] \int_{\chi_i}^{\chi_f} \mathrm{d} \chi \cdot \mathbf{F} (\chi) + {\cal O}(\gamma^2)~.
\end{equation}
Thus, if the coupling strength is weak so that it does not disturb the system too much, the leading order term of the work can be calculated from the unperturbed trajectory ensemble, in the spirit of the linear response theory.

Let us now consider applying the time-reversal operation to Eq.~\eqref{Wlinear}. First of all, the left-hand side of the equation becomes $-\langle W \rangle$, since if any amount of energy has been transferred from the system to the mechanism by the forward trajectory, we should observe the same amount of energy being transferred from the mechanism to the system for the backward trajectory. 
We can write the integral on the right-hand side in one of two ways: either considering the forward trajectory ensemble ${\cal P}[{\boldsymbol \chi}]$ while integrating the force from $\chi_f$ to $\chi_i$, or considering the backward trajectory ensemble ${\cal P}^R[{\boldsymbol \chi}]$ while integrating the force from $\chi_i$ to $\chi_f$. Choosing the second option, we  write
\begin{equation} \label{WR}
	- \langle W \rangle = \int {\cal D} {\boldsymbol \chi} ~ {\cal P}^R[{\boldsymbol \chi}] \int_{\chi_i}^{\chi_f} \mathrm{d} \chi \cdot \mathbf{F} (\chi)~.
\end{equation} 
Subtracting Eq.~\eqref{WR} from Eq.~\eqref{Wlinear} and dividing the result by 2, we get
\begin{equation} \label{WFR}
	\langle W \rangle = \int {\cal D} {\boldsymbol \chi} ~ \frac{1}{2} \left( {\cal P}[{\boldsymbol \chi}] - {\cal P}^R[{\boldsymbol \chi}] \right) \int_{\chi_i}^{\chi_f} \mathrm{d} \chi \cdot \mathbf{F} (\chi)~.
\end{equation}
This expression directly tells us that work cannot be extracted if the average entropy production measured with ${\boldsymbol \chi}$ is zero. One can see this by considering that the KL divergence Eq.~\eqref{KL} is zero if and only if the two PDFs employed to measure the divergence are identical. This means that ${\cal P}[{\boldsymbol \chi}] = {\cal P}^R[{\boldsymbol \chi}]$ if $\langle S[{\boldsymbol \chi}] \rangle = 0$, and thus $\langle W \rangle = 0$. 

If the system dynamics does not satisfy time-reversal symmetry, and the EP calculated with ${\boldsymbol \chi}$ is finite, we can connect the PDF of the forward and backward trajectories by rearranging Eq.~\eqref{S} as
\begin{equation}
	{\cal P}^R [{\boldsymbol \chi}] = {\cal P} [{\boldsymbol \chi}] e^{-S[{\boldsymbol \chi}]}~.
\end{equation}
Inserting this into Eq.~\eqref{WFR}, we get
\begin{equation}
	\langle W \rangle = \int {\cal D} {\boldsymbol \chi}\, \frac{1}{2} {\cal P}[{\boldsymbol \chi}] \left[ 1 - e^{-S[{\boldsymbol \chi}]} \right] \int_{\chi_i}^{\chi_f} \mathrm{d} \chi \cdot \mathbf{F} (\chi)
\end{equation}
Expanding the exponential, we have
\begin{equation}
	\langle W \rangle = \frac{1}{2} \int {\cal D} {\boldsymbol \chi}\, {\cal P} [{\boldsymbol \chi}] \sum_{k=1}^\infty \frac{(-1)^{k+1}}{k!} S^k[{\boldsymbol \chi}] W[{\boldsymbol \chi}] 
\end{equation}
Thus, at the linear order, the average work extracted is given as a correlation between the entropy production and the work obtained from a specific realization of a trajectory as
\begin{equation} \label{Wcor}
	\langle W \rangle \simeq \frac{1}{2} \langle S[{\boldsymbol \chi}] W[{\boldsymbol \chi}] \rangle~.
\end{equation}

In order to test Eq.~\eqref{Wcor}, we compare it to the power harnessed by the orientation-specific work extraction mechanism of our toy model. Suppose the mechanism couples to the particle when it is aligned along the $+$ direction as $s_i = +1$. As a result, the particle interacts with the work extraction mechanism when $\chi_s = (s_i, s_{i+1})$ undergoes $(+1,0) \to (0,+1)$ or $(0,+1) \to (+1,0)$ transitions. We refer to the former as the forward transition and the latter as the backward transition. Due to the weak coupling, the energy stored in the mechanism measured in units of $k_B T$ changes by $\gamma$ at each transition, with work being extracted during the forward transition and lost to the backward transition. The entropy produced during the forward transition can be calculated as $\ln(w_f / w_b) = \beta f a$, while the backward transition induces a decrease of entropy by $\ln (w_b / w_f) = -\beta f a$. Plugging these into Eq.~\eqref{Wcor} and dividing by the unit of time of our model, $\alpha^{-1}$, we estimate the power harnessed by the mechanism as
\begin{eqnarray}
    \nonumber \frac{\langle W \rangle}{\alpha^{-1}} &\simeq& \frac{\alpha}{2} \left[ (+\gamma) (\beta f a) \frac{1}{8} \frac{w_f}{\alpha + w_f + w_b } + (-\gamma) (-\beta f a) \frac{1}{8} \frac{w_b}{\alpha + w_f + w_b }  \right] \\
    &=& \frac{\gamma \beta f a }{8 } \frac{\cosh(\beta f a /2)}{1 + 2 \cosh(\beta f a/2)}~, \label{Eq:Work_toy}
\end{eqnarray}
where we have used the fact that there are a total of 8 equivalent states in the unperturbed system (4 sites $\times$ 2 orientations). 
In Fig.~4C of the main text, we compare Eq.~\eqref{Eq:Work_toy} (in solid lines) to the simulation results obtained with $\ln(w_f / w_b) = \beta f a = 0.25$ (in symbols), showing excellent agreement. The results confirm our claim that to extract finite work by weakly coupling to a given degree of freedom, the same degree of freedom should display finite EP when it is tracked in the unperturbed state.

\section{Experiment on bacteria}

\begin{figure*}[!t]
\includegraphics[width=0.6\textwidth]{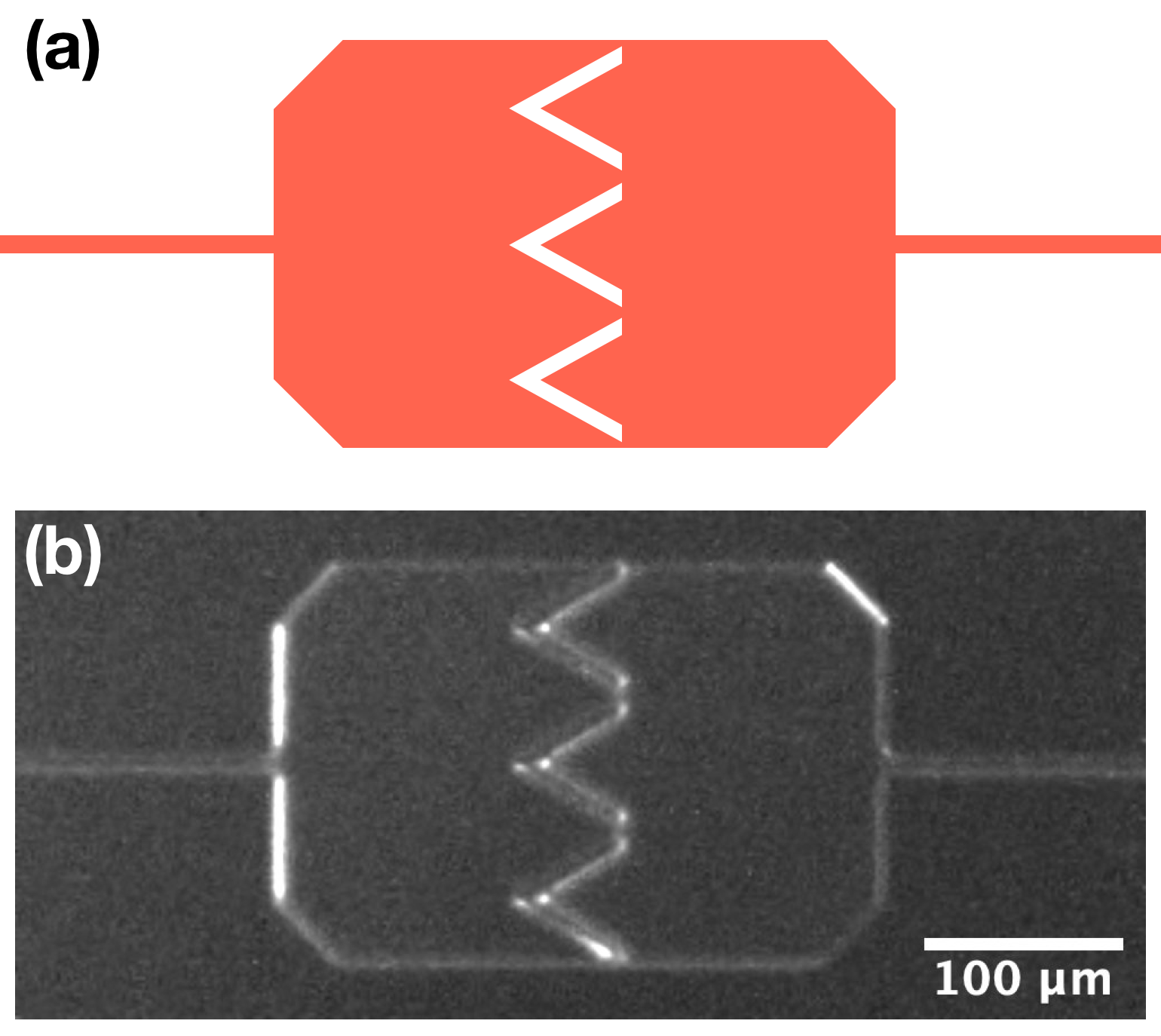} 
\caption{(A) The design of a single habitat in our ratchet device. (B) An image of a single habitat in our ratchet device under a Nikon 10$\times$ Plan APO $\lambda$ objective.} 
\label{chip_design} 
\end{figure*}

Our ratchet device is a long array consisting of many small habitats of size 200 microns $\times$ 300 microns with a row of funnels in the middle, as shown in Fig.~\ref{chip_design}. The funnels were designed with a 60$^o$ opening angle, with depth 60 microns from apex to opening and a 3 micron opening gap at the funnel vertex. 

The mask design was microfabricated, etched 10 microns deep on a 100 mm silicon wafer using standard silicon etching techniques. The wafer was then diced and 2.5 cm chips were cut out of the diced wafer. At 1000$^o$C in a O$_2$ atmosphere we grew a 0.2 micron thick SiO$_2$ layer to obtain a hydrophilic surface after the etching step. We observed that bacteria left behind a surface modification of the SiO$_2$ which prevented stable re-wetting of the chip \cite{stoney-brook1,stoney-brook2}. To enable fresh reruns the chip surfaces were cleaned by a wet HF strip of the SiO$_2$ surface followed by thermal regrowth of the SiO$_2$ surface. The top of the etched chip was sealed by pressing a 20 micron thick gas permeable LUMOX polymer film LUMOX (Sarstedt AB, N\"umbrecht, Germany) using pressurized air  (10$^4$ Pa).  Since LUMOX is highly gas permeable the bacteria in the chip where in equilibrium with 21\% O$_2$ levels at all times. All experiments were done at 20$^o$ C.

The chip was sterilized with an initial 70\% ethanol rinse followed by 2 minutes of O$_2$ plasma sterilization in a Harrick PDC-001 plasma cleaner/sterilizer (Harrick Plasma, Ithaca NY).  The plasma treatment was also important in making the chip surface temporarily hydrophilic for enhanced wetting. The sterilized chip was wet with 20 g/liter LB broth so the initial conditions are uniformly high food concentrations everywhere.

We used the bacterial K-12 strain AB1157 \cite{dewitt1962occurrence} for our experiments. To capture bacteria motion inside the ratchet device, the strain carries eGFP on the plasmid pWR21, which intrinsically express green fluorescent proteins (GFP). The bacteria were cultured in LB broth at 30$^o$C for 24 hours to reach saturation, then 20 $\mu$L of the culture was transferred into a fresh 2 mL LB broth tube for 3 hours at 30$^o$C to maintain motility under constant shaking until mid-log phase (0.6 OD). We then filled the chip with mid-log bacteria taken directly from the culture tube, and immediately covered the inoculation port with a 3 mm diameter cover glass to prevent evaporation.  

Live movies 10 minutes long at 5 frames per second were recorded through a Nikon 90i upright microscope (Nikon Instruments Inc., Melville, NY) by an Andor Neo 5.5 sCMOS camera(Andor INc., Concord, MA) using a Nikon 10$\times$ Plan APO $\lambda$ objective. These high resolution and fast rate movies allowed us to detect individual bacteria while still observing the global dynamics of the bacteria population inside the ratchet device.

\begin{figure*}[!hb]
	\includegraphics[width=0.8\textwidth]{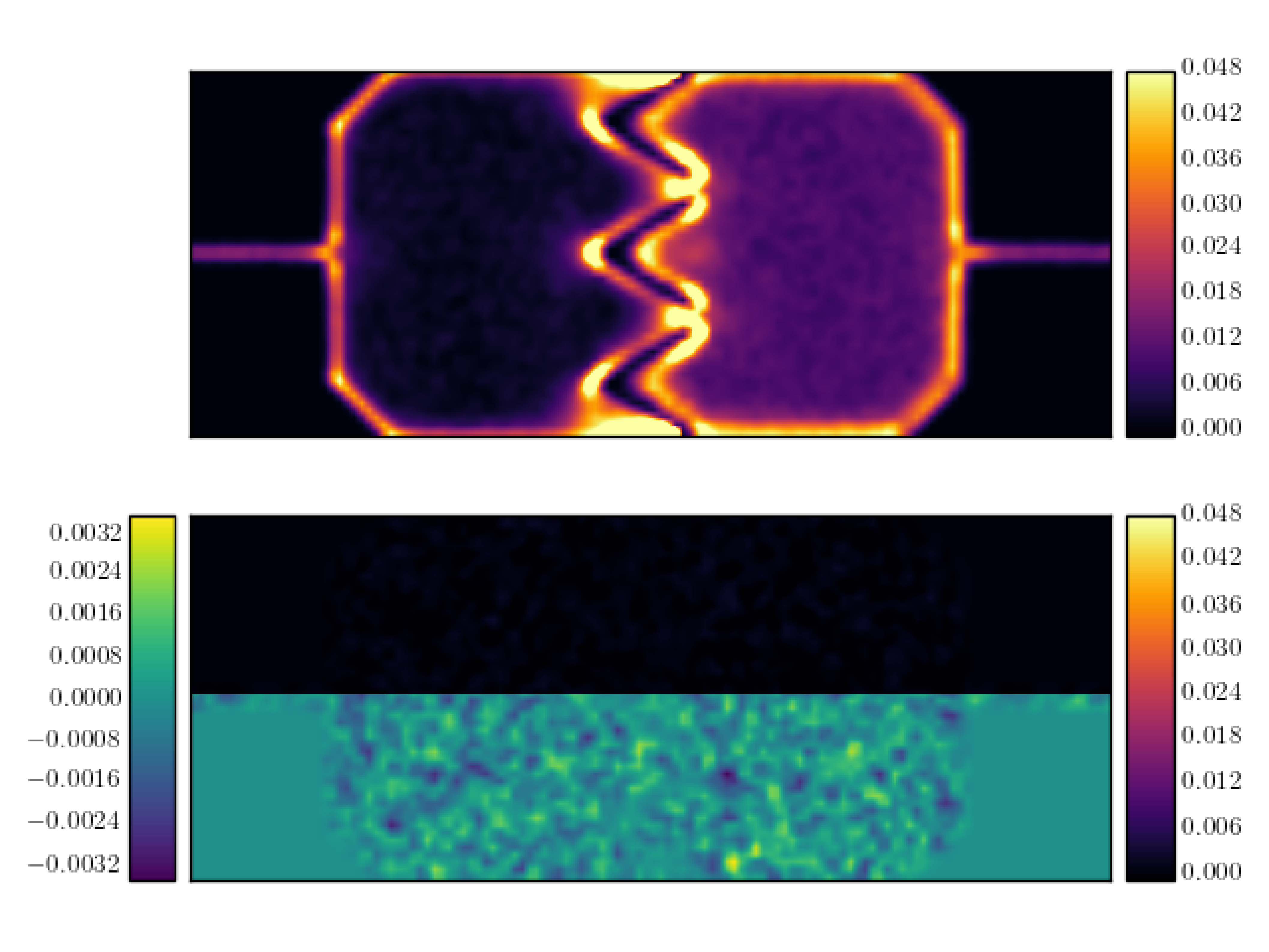} 
	\caption{ Entropy production rates of ABPs (top panel) and Brownian particles (lower panel) in a cell with funnels. The unit length of the system is the particle radius and the time step is chosen so that an ABP with a unit velocity to move a unit distance in $10^3$ steps. In the lower panel, EP is shown in two different scales - from 0 to $4.8 \times 10^{-2}$ on the upper half, and from $-3.2 \times 10^{-3}$ to $3.2 \times 10^{-3}$ for the lower half.
	}
	\label{FigS:funnel_passive} 
\end{figure*}

\section{Analysis of the bacteria experiment}

For the EP analysis of the bacterial experimental data, we discretize space using the same criteria as for the ABP simulations outlined in the main text. We track the trajectories of 2 $\times$ 2 pixel blocks and compute EP using steady state trajectories of length 17298 frames (FPS=5).

To reproduce the experimentally observed pattern of EP in simulation, we performed simulations with ABPs in a ratchet device of the same shape with $v_0=1.0$ and $D_r=0.05$. The device size is $75 \times 125$, and reflective boundary conditions are applied at the ends of the channels. 
The frames are recorded with a time interval of 1, and for spatial discretization we choose a grid size of 1.4 particle radii. We use a $2 \times 2$ block representation and the total frame number for the EP calculation is $10^6$. Note that in the simulation, we see non-zero entropy production due to collisions in the higher density region (viz. the right chamber), while in experiment we do not. This is mainly because in simulation, the system is defined strictly in 2D, so that there are actual collisions between soft disks. However, in experiment, the setup is semi-2D, so that bacteria can pass over one another and we rarely see collisions due to their excluded volume interaction.

To confirm that the observed EP is indeed due to the time-reversal symmetry breaking induced by the interaction of ABPs with the funnels, we measured EP with an identical setup but replaced ABPs with equilibrium Brownian particles. The result obtained is compared to the results for ABPs in Fig.~\ref{FigS:funnel_passive}. In the upper half of the lower panel, we plot EP for the Brownian particles on the same scale as for the map of EP obtained with ABPs (shown in the upper panel) which does not show any noticeable EP. On the lower half of the lower panel, we adjust the colorbar range to a much finer scale and observe small fluctuation around zero EP. This result confirms that it is possible to distinguish nonequilibrium dynamics from the equilibrium one using our measure.

\bibliography{bibliography}